\documentclass[11pt,tightenlines,eqsecnum,floats,aps,prd,showpacs,amsmath,superscriptaddress,amssymb,nofootinbib,longbibliography]{revtex4-1}

\usepackage{bm}
\usepackage[latin1]{inputenc}
\usepackage[spanish,english]{babel}
\usepackage{amsfonts}
\usepackage{amssymb}
\usepackage{graphicx}
\usepackage{color}
\usepackage{stmaryrd}
\usepackage{mathtools,slashed}

\usepackage{amsmath}
\usepackage{soul}
\usepackage[normalem]{ulem}

\def\d{\textrm{d}}
\newcommand{\be}{\begin{equation}}
\newcommand{\ee}{\end{equation}}
\newcommand{\bea}{\begin{eqnarray}}
\newcommand{\eea}{\end{eqnarray}}
\newcommand{\ben}{\begin{enumerate}}
\newcommand{\een}{\end{enumerate}}
\newcommand{\F}{{\rm F}}
\newcommand{\Fs}{{\prescript{\star}{}{\rm F}}}
\newcommand{\G}{{\rm G}}
\newcommand{\Gs}{{\prescript{\star}{}{\rm G}}}

\begin{document}

\title{{\bf Classical and quantum aspects of  electric-magnetic duality rotations in curved spacetimes}}

\author{Ivan Agullo}\email{agullo@lsu.edu}\affiliation{Department of Physics and Astronomy, Louisiana State University, Baton Rouge, LA 70803-4001;}
\author{Adrian del Rio}\email{adriandelrio@tecnico.ulisboa.pt}
\affiliation{Departamento de Fisica Teorica, IFIC. Centro Mixto Universitat de Valencia - CSIC.  Valencia 46100, Spain.}\affiliation{Centro de Astrof\'isica e Gravita\c{c}\~ao - CENTRA, Departamento de F\'isica, Instituto Superior T\'ecnico - IST, Universidade de Lisboa - Lisboa, Portugal;}
\author{Jose Navarro-Salas}\email{jnavarro@ific.uv.es}
\affiliation{Departamento de Fisica Teorica, IFIC. Centro Mixto Universitat de Valencia - CSIC.  Valencia 46100, Spain.}

\date{\today}

\begin{abstract}

It is well known that the source-free Maxwell equations are invariant under electric-magnetic duality rotations, $\F  \to \F \, \cos \theta+ \Fs \, \sin \theta $. These transformations are indeed a symmetry of the theory in Noether sense. The associated  constant of motion is the difference in the intensity between self- and anti-self dual components of the electromagnetic field or, equivalently, the difference  between the right and left circularly polarized components. 
This conservation law holds  even if the electromagnetic field interacts with an arbitrary classical gravitational background. 
After re-examining  these results,  we discuss whether this symmetry is maintained 
when the electromagnetic field is quantized. 
The answer is in the affirmative in the absence of gravity, but not necessarily otherwise. 
As a consequence, the net polarization of the quantum electromagnetic field  fails to be conserved in curved spacetimes.  This is a quantum effect, and it can be understood as the  generalization of the fermion chiral anomaly to  fields of spin one.

\end{abstract}


\maketitle

\section{  Introduction } \label{Introduction}

Symmetries play an important  role in many areas of science. They  
are widely considered as guiding principles for constructing  physical theories, and their connection with conservation laws 
 found by Noether  
one century ago \cite{Noetherpaper}  is a cornerstone of modern physics. An interesting example is given by Maxwell's theory of electrodynamics, whose  invariance under  Poincaré transformations leads to conservation of energy, linear and angular momentum. (The invariance extends   in fact  to the full conformal group.)
The theory is also invariant under gauge transformations when the electromagnetic potential is introduced, and when it is coupled to matter fields the symmetry is related to  
the conservation of electric charge. Furthermore, in the absence of charges and currents, this theory enjoys a peculiar symmetry (in four spacetime dimensions).
It is a simple exercise to check that  Maxwell's equations, and also the stress-energy tensor, are invariant under the `exchange' of the electric and magnetic fields  $\vec E \to \vec B$, $\vec B \to - \vec E$, as first noticed  in the early years  after the introduction of Maxwell equations.  
 This discrete, $\mathbb Z_2$  operation is commonly known as a  duality transformation. But the invariance of Maxwell's equations extends to  $SO(2)$  rotations $\vec E \to \vec E \, \cos \theta +\vec B \, \sin \theta$, $\vec B \to \vec B \, \cos \theta -\vec E \, \sin \theta$, of which the duality transformation is just the particular case with $\theta =\pi/2$.
Although apparently innocuous,  
this {\it continuous} transformation has revealed in more recent times to have interesting consequences.

 In the mid sixties, Calkin  pointed out that these  transformations leave  Maxwell's action invariant, and identified the associated conserved charge  as the difference  between the intensity of the right- and left-handed circularly polarized components of the electromagnetic field \cite{Calkin1965}. This conservation law was studied in more detail by  Deser and Teitelboim  in \cite{DeserTeitelboim1976, Deser1982},  and {\em proved to remain true in curved spacetimes}. This quantity is sometimes known as the optical helicity \cite{BarnettCameronYao2012}, and it  also agrees  with   the V-Stokes parameter. Henceforth, besides conservation of energy and momentum, the polarization of electromagnetic radiation will be also a constant of motion as long as no electromagnetic sources are present, courtesy of the  symmetry under electric-magnetic  rotations.

A natural question now is to analyze whether this symmetry continues to hold in {\em quantum} electrodynamics. If $j^{\mu}_D$ is the  Noether current associated with electric-magnetic rotations, this task reduces to check if the vacuum expectation value $\left< \nabla_{\mu}j^{\mu}_D\right>$ vanishes. In contrast to the classical theory, this is a non-trivial calculation, that involves appropriate renormalization of ultraviolet divergences. It is well-known that quantum fluctuations produce off-shell contributions to physical quantities that might spoil  classical symmetries. When this occurs, we speak of a quantum anomaly in the theory.

Historically, the issue of quantum anomalies  first appeared in the seminal works by Adler, Bell and Jackiw, as a result of solving the pion decay puzzle \cite{Adler1969, BellJackiw1969}. They found that the  chiral symmetry of the action of a massless Dirac field breaks down at the quantum level when the  fermionic field interacts with  an electromagnetic   background. Namely, they obtained  the celebrated chiral or axial anomaly $\left< \nabla_{\mu}j^{\mu}_A\right>= -\frac{\hbar q^2}{8\pi^2} \F_{\mu\nu}\Fs^{\mu\nu}$, where $j^{\mu}_A$ is the fermionic chiral current, $\F_{\mu\nu}$ the field strength of the background electromagnetic field, $\Fs_{\mu\nu}$ its dual,  and $q$ the charge of the fermion. Soon after,  a similar anomaly was found when the massless Dirac field is immersed in a classical gravitational background  \cite{Kimura1969},  \cite{DelbourgoSalam1972},  \cite{EguchiFreund1976}, $\left< \nabla_{\mu}j^{\mu}_A\right>= \frac{\hbar}{192\pi^2} R_{\mu\nu\alpha\beta}\, ^{\star}R^{\mu\nu\alpha\beta}$, where $R_{\mu\nu\alpha\beta}$ is the Riemann tensor. These discoveries led to an outbreak of interest in anomalies both in quantum field theory and  mathematical physics, leading to further  examples and a connection with the well-known index-theorems in geometric analysis \cite{EguchiHansonGilkey1980, BastianelliVanNieuwenhuizen2006, Nakahara2003}. The existence of anomalies has important physical  implications. Besides the prediction of the neutral pion decay rate to two photons, they have applications in studies of the matter-antimatter asymmetry of the universe, the U(1) and strong CP problems in QCD, and provide a deeper understanding of the Standard Model via anomaly cancelation \cite{Schwartz2014}. These cancelations  have  played a major role in string theories and supergravity too (for a detailed account see, for instance,  \cite{BastianelliVanNieuwenhuizen2006} and references therein).
A decade later of the discovery of the chiral anomaly, the nature of quantum anomalies was further clarified by Fujikawa, using the  language of path-integrals \cite{Fujikawa1979, Fujikawa1980}. 
He found that the existence of anomalies can also be understood as the failure of  the measure of the path integral to respect the symmetries of the action.  Fujikawa's arguments provided an alternative and elegant way of computing anomalies. 

In this paper, we prove that electric-magnetic rotations are also anomalous, provided the electromagnetic field propagates in a sufficiently non-trivial spacetime. To meet our goal, we write Maxwell's theory in terms of self- and anti self-dual variables, which will make the structure of the   theory  significantly more transparent, particularly in the absence of charges and currents. In fact, in these variables duality rotations look mathematically---and physically---similar to chiral transformations of massless spin $1/2$ Dirac fields, and in this sense our result can be understood as the spin $1$ generalization of the fermionic  chiral anomaly.  We derive our result by using two complementary methods, namely by directly computing $\left<\nabla_{\mu} j_D^{\mu}\right>$ using the method of heat kernel-renormalization, and by Fujikawa's path-integral approach.

This paper is organized as follows. In Sec.\ II we review the analysis of the classical duality symmetry in source-free electrodynamics, and derive the  associated  Noether charge and current. We do it both in the Lagrangian and Hamiltonian frameworks. In Sec.\ III we introduce self- and anti self-dual variables, and emphasize their advantages in the source-free theory. We will show how Maxwell's equations can be conveniently written as first order equations, either for fields or potentials, that are analog to    Weyl's equations for spin $1/2$ fields.   Sec.\ IV will derive a first-order action for Maxwell electrodynamics in  self- and anti self-dual variables, which makes the theory formally analog to  Dirac's theory of massless fermions. 
Sec.\ V  deals with the quantum theory, and the derivation of the quantum electromagnetic duality anomaly, by using the two methods mentioned above.  We finally give some concluding remarks in Sec. VI. To alleviate the main text of the article, we have moved many of the mathematical details and calculations to appendices A-G. %

A shorter version of this work  appeared in \cite{AgullodelRioPepe2017}. Here we provide  further details,  alternative avenues of arriving to the final result, and correct some minor errors which translate into a different numerical factor in the result for $\left<\nabla_{\mu} j_D^{\mu}\right>$.

We  follow the convention $\epsilon^{0123}={1/\sqrt{-g}}$ and  metric signature $(+, -, -, -)$.   More specifically, we follow the $(-,-,-)$ convention of \cite{MTW1973}. We   restrict to 4-dimensional spacetimes and  assume the Levi-Civita connection.  We use Greek indices $\mu,\nu,\alpha,\cdots$ for tensors in curved spacetimes, while latin indices $a,b,c,\cdots$ are used for tensors in  Minkowski spacetime. Indices $I, J,K,\cdots$ or $\dot I, \dot J,\dot K,\cdots$ refer to tensors in an internal  space associated with the spin $1$  complex Lorentz representations. Unless otherwise stated, we assume all fields to be smooth and to have standard fall-off conditions at infinity. We use  units for which $c=1$.

\section{Classical theory and electric-magnetic rotations}\label{sec:2}

\subsection{Lagrangian formalism} \label{Lagrangian formalism}

In this paper we are concerned with free Maxwell's theory, i.e.\ electromagnetic fields in the absence of electric charges and currents, formulated on a globally hyperbolic spacetime $(M, g_{\mu\nu})$ with metric tensor $g_{\mu\nu}$. The classical theory is described by the action  
\be \label{Maction} S[A_{\mu}]=-\frac{1}{4}\int \d^4 x\sqrt{-g}  \, \F^{\mu\nu}\F_{\mu\nu} \, \ee
where $\F$ is a closed two-form ($\d\F=0$) defined in terms of its potential $A$ as $\F=  \d A$, or more explicitly, $\F_{\mu\nu} = \nabla_{\mu} A_{\nu}-\nabla_{\nu} A_{\mu}$. Maxwell's equations  read $\Box A_{\nu}-\nabla^{\mu}\nabla_{\nu}A_{\mu}=0$, where $\nabla$ is the covariant derivative associated with $g_{\mu\nu}$ and $\Box\equiv g^{\mu\nu} \nabla_{\mu}\nabla_{\nu}$. When written in terms of the dual tensor $\Fs$, these equations take the compact form $\d \Fs=0$ and, together with $\d\F=0$, make manifest that  the field equations are invariant under electric-magnetic rotations
\bea \label{dt} \F &\longrightarrow& \F \, \cos \theta+ \Fs \, \sin \theta \, , \nonumber \\
\Fs &\longrightarrow& \Fs \, \cos \theta- \F \, \sin \theta \, .\eea
For $\theta =\pi/2$ one has the more familiar duality transformation  $\F \to \Fs$  and $\Fs \to -\F$.  If this one-parameter family of transformations are a true symmetry of the action, then Noether's analysis must provide a conserved charge associated to it. We now analyze this problem. Our presentation simply re-phrases in a manifestly covariant way the results of Ref.\ \cite{DeserTeitelboim1976}.

For the transformation (\ref{dt}) to be a symmetry, its infinitesimal version $(\delta \F=\Fs \, \delta \theta)$ must  leave the action invariant or, equivalently, the Lagrangian density $ \mathcal{L}=-1/4 \sqrt{-g} \, \F_{\mu\nu}\F^{\mu\nu}$ must change by a total derivative, $\delta \mathcal{L}=\sqrt{-g}\, \nabla_{\mu} {h^{\mu}}$, for some current $h^{\mu}$. This must be true even off-shell, i.e.\ when $\F$ and $\Fs$ do not satisfy the equations of motion. In analyzing if this is the case one faces two issues. On the one hand,  since $\F$ is  a closed two-form (i.e.\ $\d\F=0$), for the transformation $\delta \F=\Fs \, \delta \theta$ to be consistent  $\Fs$ must be also closed; but this amounts to say that equations of motion hold. In other words, the transformation (\ref{dt}) can only be consistently defined 
on-shell.
\footnote{This ``difficulty''  is singular of the second order formalism. If one uses a first order Lagrangian, or a Hamiltonian formulation, the usual electric-magnetic rotations can be implemented off-shell. This point has been emphasized in \cite{Deser1982}, and will be made explicit in the next subsection and in section \ref{First order Lagrangian formalism: Dirac-type formulation}.}  And secondly,  since the usual configuration variables of Maxwell's action are the vector potential $A$ rather than the field $\F$, to apply Noether's techniques we first need to re-write (\ref{dt}) in terms of $A$. A convenient strategy  to deal with these two issues  is to define a more general transformation, that will agree with electric-magnetic rotations only on-shell, as follows
\be \label{dtA} \delta A_{\mu}=Z_{\mu}\, \delta \theta \, ,\ee 
where $Z_{\mu}$ is  implicitly defined by $\d Z\equiv \Fs+\G$, and $\G$ is a two-form that is subject to the following conditions, but arbitrary otherwise: 
\begin{enumerate}
\item $\G$ vanishes only when $A_{\mu}$ satisfies the equations of motion, $\G|_{\rm on-shell}=0$. 
 This ensures that $\d Z=\Fs$ on-shell, and then (\ref{dtA}) reduces to the usual electric-magnetic transformation.

\item $\G$ is not closed, $\d \G\neq 0$---unless the equations of motion hold. This guarantees that $\Fs$ is not  closed (off-shell).

\item $\G$ has zero magnetic part relative to {an arbitrary} observer, i.e. $n^{\nu} \, {\Gs}_{\mu\nu}=0$, where $n^{\nu}$ is a time-like vector field, and $\Gs$ is the dual of $\G$. (This condition is  equivalent  to say that the electric field relative to the observer satisfies Gauss's law.)

\end{enumerate}

Note that $Z_{\mu}$ is a non-local functional of $A_{\mu}$. 
However, as discussed in \cite{DeserTeitelboim1976}, this is not an impediment to apply Noether's formalism. 

Under the transformation (\ref{dtA}), we obtain (see Appendix \ref{Noether current} for more details)
\be
\delta \mathcal L =  - \delta \theta \, \frac{\sqrt{-g}}{2}\, \nabla_{\mu}\left[  A_{\nu}{\Fs}^{\mu\nu}-Z_{\nu}\, {(\d Z)}^{\mu\nu}\right]  \equiv  \sqrt{-g}\, \nabla_{\mu} h^{\mu} \, ,\label{variacionlagrangiano} 
\ee
confirming that  electric-magnetic rotations are  a symmetry of  source-free Maxwell's theory. The conserved Noether  current $j_D^{\mu}$ associated with this symmetry is
\be
j_D^{\mu} = \frac{1}{\sqrt{-g}} \,   \frac{\partial \mathcal L}{\partial\nabla_{\mu} A_{\nu}}\delta A_{\nu}-h^{\mu}=  \frac{ 1}{2}\,  \Big[  A_{\nu}{\Fs}^{\mu\nu}-Z_{\nu}\,  \F^{\mu\nu}-Z_{\nu} \, \Gs^{\mu\nu}\Big] \, ,\label{jND0}
\ee
(we have dropped $\delta\theta$ from the definition of $j_D^{\mu} $). This current is gauge-dependent. But this is not a problem either, as long as the associated conserved  charge is gauge invariant, which is in fact the case. When evaluated on-shell (i.e.\ when  $\G=0$, and therefore $\d Z=\Fs$)
\be \label{osj} j_D^{\mu}|_{\rm on-shell} = \frac{ 1}{2}\,   \Big[  A_{\nu}\, {\Fs}^{\mu\nu}-Z_{\nu}\,  \F^{\mu\nu}\Big] \, . \ee
Now, if we foliate the spacetime using a one parameter family of Cauchy hyper-surfaces $\Sigma_t$, the quantity
\be \label{Cargadual} Q_D=\int_{\Sigma_t} \d \Sigma_{\mu}   \, j_D^{\mu} =- \frac{1}{2}\,  \int_{\Sigma_t} \d\Sigma_3 \,  \left(A_{\mu}B^{\mu}-Z_{\mu}E^{\mu}\right) \, , \, \ee
is a conserved charge, in the sense that it is independent of the choice of ``leaf'' $\Sigma_t$.  In this expression, $\d\Sigma_3$ is the volume element in $\Sigma_t$, and $E^{\mu}=n_{\nu}\F^{\mu\nu}$ and $B^{\mu}=n_{\nu}\Fs^{\mu\nu}$ are the electric and magnetic parts,  respectively, of the electromagnetic tensor  field $\F$ relative to the foliation $\Sigma_t$. 
The same expression for $Q_D$ is obtained if  $j_D^{\mu}|_{\rm on-shell}$ is used in  place of $j_D^{\mu}$ in (\ref{Cargadual}), and hence the conserved charge is insensitive to the extension of the transformation done above by the introduction of  $\G$.

 One can check, by explicit computation, that  $\nabla_{\mu} j_D^{\mu}= -Z_{\nu} \nabla_{\mu}F^{\mu\nu}$, and therefore $\nabla_{\mu} j_D^{\mu}=0$ when the equations of motion $\nabla_{\mu}F^{\mu\nu}=0$ hold. In the quantum theory, however, off-shell contributions of quantum origin  may spoil the conservation of the current. 
The calculation of the expectation value of $\nabla_{\mu} j_D^{\mu}$ using the formalism derived in this section is complicated, since it would involve the operator $Z_{\mu}$, which is a (non-local) functional of the configuration variable $A_{\mu}$.\footnote{The first term in (\ref{jND0}) and its quantum aspects have been previously discussed in \cite{Dolgovetal1989} (see also \cite{Reuter1988}). However, this term by itself is not conserved classically (something that  cannot be fixed by any gauge transformation), and in fact its associated ``charge"  does not generate duality rotations in phase space (see Sec. IIB). Therefore, the first term in (\ref{jND0})  alone is not associated with the symmetry under electric-magnetic rotations. The fact that its vacuum expectation value does not vanish, although of  physical interest in its own right,  does not really prove the existence of an anomaly, as claimed in \cite{Dolgovetal1989}. Other vacuum expectation values of physical interest have  been computed in \cite{AgulloLandetePepe}.}  
 This difficulty can be alleviated by working in phase space, since there one can treat $Z_{\mu}$ and $A_{\mu}$ as independent fields. This motivates the Hamiltonian analysis of the next  subsection, and the use of a first order formalism in the rest of the paper. In particular, in sections  \ref{Electrodynamics in terms of self- and anti self-dual variables} and  \ref{First order Lagrangian formalism: Dirac-type formulation} we will re-derive  $j^{\mu}_D$ in a first-order Lagrangian formalism  using  self- anti self-dual variables. This will make  the derivation significantly  more transparent. The physical interpretation of $Q_D$ will become also more clear, and  we postpone the discussion until then. \\

\subsection{Hamiltonian formalism} \label{Hamiltonian formalism}

The Hamiltonian formalism provides a complementary approach to the study of the electric-magnetic symmetry, and in this subsection we briefly summarize the derivation of $Q_D$ {following} this framework. 
We will restrict here to Minkowski spacetime, since the generalization to curved geometries  using the standard vector potential and  electric field  as canonical coordinates  becomes cumbersome. 

Given an inertial frame in Minkowski spacetime, Maxwell's Lagrangian (\ref{Maction}) takes the form
\be L=\int_{\mathbb R^3} \d^3x \, \mathcal L=\int_{\mathbb R^3} \d^3x \, \frac{1}{2}\, \Big[(\dot {\vec A}-\vec \nabla A_0)^2-(\vec \nabla \times \vec A)^2\Big] \, ,\ee
 where $\vec \nabla$ is the usual three-dimensional derivative operator.  Our conventions are $\vec A\equiv (A_1,A_2,A_3)$,  $\vec E\equiv (E_1,E_2,E_3)$,  $E_i\equiv F_{i0}$, and $\vec E^2\equiv E_1^2+E_2^2+E_3^2$. From this, we see that the canonically conjugate variable of $\vec A$ is the electric field $\frac{\delta L}{\delta \dot{A}_i}=E^i$, and the conjugate variable of $A_0$ vanishes, since the Lagrangian does not involve $\dot A_0$. Then,  $A_0$ is a Lagrange multiplier, and from its equation of motion one obtains a constraint, the familiar Gauss' law  $\vec \nabla \cdot \vec E=0$. Then, the canonical phase space is made of pairs $(\vec A(\vec{x}), \vec E(\vec{x}))$, with a symplectic, or  Poisson structure given by $\{A_i(\vec{x}),  E^j(\vec{x}')\}=\delta_i^j \, \delta^{(3)}(\vec{x}-\vec{x'})$. A Legendre transformation produces the Hamiltonian 
\be H=\int_{\mathbb R^3} \d^3x \, \frac{1}{2}\, \Big[\vec{E}^2 + (\vec \nabla \times \vec A)^2-A_0\,  (\vec \nabla \cdot \vec E)\Big]\, ,\ee
where we have disregarded a boundary term. In Dirac's terminology, $\vec \nabla \cdot \vec E=0$ is a first class constraint, and  it tells us that there is a gauge freedom in the theory, given precisely by the canonical transformations generated by $\vec \nabla \cdot \vec E$. 

Hamilton's equations read
\bea \dot{\vec A}&=&\{\vec A,H\}=-\vec E-\vec \nabla A_0 \, \nonumber \\ \dot{\vec E}&=&\{\vec E,H\}=\vec \nabla \times (\vec \nabla \times \vec A)\, . \label{HamEQ} \eea
where $A_0(\vec{x})$ is now interpreted as an arbitrary function without dynamics, and the term proportional to it in the expression for $\dot{\vec A}$ corresponds precisely to the gauge flow. These six equations, together with the Gauss constraint, are equivalent to standard Maxwell's equations (once we define $\vec B\equiv \vec \nabla \times \vec A$).\\

Electric-magnetic rotations in phase space are given by 
\bea \label{dtps2} \delta \vec E=(\vec \nabla \times \vec A)\equiv \vec B\, , \hspace{1cm} \delta \vec A&=&-(\vec \nabla \times)^{-1} \vec E\equiv \vec Z\,  , \eea
 where $(\vec \nabla \times)^{-1}$  is the inverse of the curl; when acting on  traverse fields---such as $\vec E$---it can be easily computed by using the relation $(\vec \nabla\times)^{-1}=-\nabla^{-2}\, \vec \nabla \times$. The presence of the operator $(\vec \nabla\times)^{-1}$ in (\ref{dtps2}) makes evident that we are dealing with a transformation that is non-local in space.  
 
 Now, the  generator of the transformation  (\ref{dtps2}) can be easily obtained by computing the symplectic product of $(\vec A, \vec E)$ and $(\delta \vec A, \delta \vec E)$:
 \be \label{Cargadual2} Q_D=\Omega[(\vec A, \vec E),(\delta \vec A, \delta \vec E)]=- \frac{1}{2}\int_{\mathbb R^3} d^3x\, [\vec E\cdot \delta \vec A-\vec A\cdot  \delta \vec E]=\frac{1}{2}\int_{\mathbb R^3} d^3x\, [\vec A\cdot  \vec B- \vec Z \cdot \vec E]\, . \ee
$Q_D$ is independent of $A_0$, and by integrating by parts it is easy to show that only the transverse part of $\vec A$ and $\vec Z$ contribute to $Q_D$; hence it is gauge invariant. It is also  straightforward to check that $Q_D$ is indeed the correct generator, since $\delta \vec A=\{\vec A,Q_D\}$ and $\delta \vec E=\{\vec E,Q_D\}$ reproduce expressions (\ref{dtps2}).
To finish, one can now check  that $\dot Q_D=\{Q_D, H\}=0$. Therefore, $Q_D$ is a constant of motion. This implies that   the canonical transformation generated by  $Q_D$ is a symmetry of the theory.

\section{Electrodynamics in terms of self- and anti self-dual variables} \label{Electrodynamics in terms of self- and anti self-dual variables}

Many  aspects of   Maxwell's theory  in absence of charges and currents become more transparent when self- and anti self-dual variables are used (see e.g.\ \cite {Weinberg1964, DowkerDowker1966, Penrose1965}). Some of the advantages of these variables are well-known and, in particular, they are commonly used in the spinorial formulation of electrodynamics \cite{PenroseRindler1984}. 
For the sake of clarity, we introduce these variables first in Minkowski spacetime, and  extend the formalism later to curved geometries. 

\subsection{Minkowski spacetime} \label{Minkowski spacetime}
The self- and anti self-dual components of of the electromagnetic field are defined as $ \vec H_{\pm}\equiv\frac{1}{\sqrt{2}} \, (\vec E\pm i\, \vec B)$. We now enumerate the properties and interesting aspects of these complex variables.

\begin{enumerate} 
\item {\bf Electric-magnetic rotations}

The transformation rule of the electric and magnetic fields  under electric-magnetic rotations
\bea \vec E &\longrightarrow& \vec E \, \cos \theta +\vec B \, \sin \theta \, , \nonumber \\
 \vec B &\longrightarrow& \vec B \, \cos \theta -\vec E \, \sin \theta \, , \eea
translates to   
\be \vec H_{\pm} \longrightarrow  e^{\mp \, i \theta} \vec H_{\pm}\, . \ee
An ordinary duality transformation $ \vec E\to\vec B$, $ \vec B\to-\vec E$ 
corresponds to $\theta=\pi/2$. 
Then, this operation produces\footnote{It is common to add the imaginary unit  $i$ because in that way this operation  has real eigenvalues, and it can be represented by a self-adjoint operator in the quantum theory.} $i\, \vec H_{\pm} \to \pm\, \vec H_{\pm}$. It is for this reason that $\vec H_+$ and $\vec H_-$ are called the self- and anti self-dual components of the electromagnetic field, respectively. 

\item {\bf Lorentz transformations}

The components of  $\vec E$ and $\vec B$ mix with each other under a Lorentz transformation. For instance, under a boost of velocity $v$ in the $x$-direction
\bea \vec E=(E_x,E_y,E_z)&\longrightarrow& [E_x,\gamma\, (E_y-v\, B_z), \gamma \, (E_z+v\, B_y)] \, , \nonumber \\
 \vec B=(B_x,B_y,B_z)&\longrightarrow& [B_x,\gamma\, (B_y+v\, E_z), \gamma \, (B_z-v\, {E_y}) ] \, ,\eea
 where $\gamma=1/\sqrt{1-v^2}$. This  transformation does not correspond to any irreducible representation of the Lorentz group. However, when $\vec E$ and $\vec B$ are combined into $\vec H_{\pm}$, it is easy to see that the components of $\vec H_+$ and $\vec H_-$ no longer mix
\be \vec H_{\pm}=(H_{\pm}^x,H_{\pm}^y,H_{\pm}^z)\longrightarrow [H_{\pm}^x,\gamma\, (H_{\pm}^y\pm i\,  v\, H_{\pm}^z), \gamma\, (H_{\pm}^z\mp i\, \, v\, H_{\pm}^y)] \, . \ee
These are the transformation rules associated with the two irreducible representations of the Lorentz group for fields of spin $s=1$. They are the so-called $(0,1)$ representation for  $\vec H_+$, and the $(1,0)$ one for $\vec H_-$. More generally, for any element of the  restricted Lorentz group $SO^+(1,3)$ (rotations + boots), the infinitesimal transformation reads
\be  H_{\pm}^J  \longrightarrow  [D(\epsilon_{ab})]_{IJ}\, H_{\pm}^J =\Big[\delta_{IJ}-\,\frac{1}{2} \epsilon_{ab}\, ^{\pm}\Sigma^{ab}_{\ \  IJ}\Big]\, H_{\pm}^J  \, ,\ee
(upper case latin indices $I,J,K,\cdots$ take values from 1 to 3) where {$\delta_{IJ}$ is the Kronecker delta}, $ ^{\pm}\Sigma^{ab}_{\ \  IJ}$ are the generators of the $(0,1)$ and $(1,0)$ representations\footnote{They satisfy the algebra $[^{\pm}\Sigma^{a b} , ^{\pm}\Sigma^{c d} ] = (\eta^{ac}  \, ^{\pm}\Sigma^{b d}-\eta^{ad}\, ^{\pm}\Sigma^{b c}+\eta^{bd}  \, ^{\pm}\Sigma^{ac}-\eta^{bc}  \, ^{\pm}\Sigma^{a d})$.}, and the anti-symmetric matrix $\epsilon_{ab}=\epsilon_{[ab]}$ contains the parameters of the transformation.  The use of self- and anti self-dual fields $ \vec H_{\pm}$ makes more transparent the fact that electrodynamics describes massless fields of spin $s=1$. 

\item {\bf Maxwell's equations}

The equations of motions for $\vec E$ and $\vec B$, 
\bea \label{Meq} \vec \nabla \cdot \vec E&=&0\, , \hspace{2cm} \vec \nabla \cdot \vec B =0 \; , \nonumber \\
\vec \nabla \times \vec E &=&-\partial_t\, \vec B \, , \hspace{1cm} \vec \nabla \times \vec B =\partial_t\, \vec E\, .
\eea
when written in terms of $\vec H_{\pm}$, take the form
\be \label{HMeq} \vec \nabla \cdot \vec H_{\pm}=0\, , \hspace{1cm} \vec \nabla \times \vec H_{\pm} =\pm i\, \partial_t\, \vec H_{\pm} \, .
\ee
Notice that, in contrast to $\vec E$ and $\vec B$, the self- and anti self-dual fields are not coupled by the dynamics. The equations for $\vec H_-$ and $\vec H_+$ are related by  complex conjugation. 

Equations (\ref{HMeq}) are  linear, and  therefore the space of solutions has structure of vector space. It is spanned by positive- and negative-frequency solutions:
\be \label{solH} \vec H_{\pm}(t,\vec x)=\int_{\mathbb R^3} \frac{\d^3 k}{(2\pi)^3} \, \Big[ h_{\pm}(\vec k)\, e^{-i (k\, t-\vec k\cdot \vec x)}+ \bar h_{\mp}(\vec k)\, e^{i (k\, t-\vec k\cdot \vec x)} \Big] \, \hat \epsilon_{\pm}(\vec k)\, , \ee
where $k=|{\vec k}|$, and  $h_{\pm}(\vec k)$ are complex numbers that indicate the amplitude of the positive and negative frequency components of a particular solution (the ``bar" denotes complex conjugation). The polarization vectors are given by $\hat \epsilon_{\pm}=\frac{1}{\sqrt{2}}\, (\hat e_1\pm i\, \hat e_2)$ where  $\hat e_1(\vec k)$ and $\hat e_2(\vec k)$ are  two unit-vectors that, together with $\hat k$, form an orthonormal triad of space-like vectors, with orientation  $\hat e_1 \times \hat e_2=\hat k$. 

The explicitly  form  (\ref{solH}) of a generic solution  helps to understand the relation between self- or anti self-duality and helicity in Minkowski spacetime. By paying attention to the way the electric and magnetic parts (i.e.\ the real and imaginary parts of $\vec H_{\pm}$, respectively) rotate with respect to the direction of propagation $\hat k$ during the course of time, one finds the following relation:
\begin{itemize}
\item Positive-frequency Fourier modes  $e^{-i (k\, t-\vec k\cdot \vec x)}\, \hat \epsilon_{\pm}(\vec k)$ have positive helicity (that corresponds to left-handed circular polarization) for self-dual fields, and negative helicity for anti self-dual fields.
\item For negative-frequency modes  $e^{i (k\, t-\vec k\cdot \vec x)}\, \hat \epsilon_{\pm}(\vec k)$  the relation is inverse: they have negative helicity (right-handed circular polarization) for self-dual fields, and positive helicity for anti self-dual fields.
\end{itemize}
We see that duality and helicity are closely related concepts in Minkowski spacetime, although the relation is not trivial; one needs to distinguish between self- and anti self-dual fields {\em and}  positive and negative frequencies \cite{Ashtekar:1986ec}. This is the analog of the familiar relation between chirality and helicity for massless spin $1/2$ fermions. In this sense, duality is the chirality of photons. 

Furthermore, in more general spacetimes where neither Fourier modes nor the notion of positive and negative frequency are anymore useful,  self- or anti self-duality generalizes the concept of helicity, or handedness of electromagnetic waves.

\item {\bf Self- and anti self-dual potentials}

The constraints $\vec \nabla \cdot \vec H_{\pm}=0$ allow us to define the potentials  $ \vec A_{\pm}$ by:
\be \label{defApm} \vec H_{\pm}=\pm \, i\, \vec \nabla \times \vec A_\pm  . \ee
It is clear from this definition that the longitudinal part of $\vec A_\pm$ contains a gauge ambiguity consisting in adding the divergence of an arbitrary scalar function. 
Note also that no time derivatives have been involved in the definition of these potentials.

\item {\bf Maxwell's equations for  potentials}

Substituting  (\ref{defApm}) in the field  equations (\ref{HMeq}), produces 
\be \label{AMeq} \pm\, i \vec \nabla \times \vec A_{\pm} = -\, \partial_t\, \vec A_{\pm} + \vec \nabla A^0_{\pm} \, .\ee
These equations by themselves are equivalent to Maxwell's equations. It may be surprising at first that Maxwell's theory can be written as first order equations for  potentials. The reason comes from the fact that in---and only in---the source-free theory,  in addition to the standard potential $\vec A$ defined from $\vec B=\vec \nabla\times \vec A$, Gauss's law $\vec \nabla \cdot \vec E=0$ allows us to define a second potential $\vec Z$, as $\vec E\equiv-\vec \nabla\times \vec Z$. Then, the first order equations 
\bea 
\dot{\vec A}&=&\vec \nabla \times \vec Z+\vec \nabla A_0 \, , \nonumber \\
\dot{\vec Z}&=&-\vec \nabla \times \vec A+\vec \nabla Z_0 \, ,  \eea
are equivalent to  Maxwell equations (to see this, take curl and use the relation between potentials and fields). Therefore,  Maxwell's equations can be written as first order equations for  potentials at the expenses of duplicating the number of potentials. The relation between the two sets of potentials is $A_a^{\pm}=\frac{1}{\sqrt{2}}(A_{a}\pm\, i\,  Z_{a})$.

\item{\bf Manifestly Lorentz-covariant equations}

The equations (\ref{HMeq}) and (\ref{AMeq}) for fields and potentials can be re-written in a more compact way as
\be \label{1oeq}  \alpha^{ab}_{I} \partial_{a} H^I_+=0\, , \hspace{1cm} \bar \alpha^{ab}_{I} \partial_{a} A^{+}_{b}=0 \, . \ee
The equations for $ H_-$ and $A_{-\, }$ are obtained by complex conjugation. In these expressions $\alpha^{ab}_{I} $ are three $4\times4$ matrices, for $I=1,2,3$, and the bar over $\alpha^{ab}_{I}$ indicates complex conjugation. The components of these matrices in an inertial frame can be identified by comparing these equations with  (\ref{HMeq}) and (\ref{AMeq}): 
 \bea \label{alphas}
\alpha^{ab}_1 =   \left( {\begin{array}{cccc}
  0 &- 1 & 0& 0  \\
1 & 0 & 0 & 0  \\
0 & 0 & 0 & i  \\
0 & 0 & -i & 0  
  \end{array} } \right)  \, ,  
\hspace{0.3cm} 
\alpha^{ab}_2  = \left( {\begin{array}{cccc}
  0 & 0 & -1& 0  \\
0 & 0 & 0 & -i  \\
1 & 0 & 0 & 0 \\
0 & i & 0 & 0  
  \end{array} } \right) \, ,
  \hspace{0.3cm}
  \alpha^{ab}_3  =   \left( {\begin{array}{cccc}
  0 & 0 & 0& -1  \\
0 & 0 & i & 0  \\
0 & -i & 0 & 0 \\
1 & 0 & 0 & 0  
  \end{array} } \right)  \, .
  \eea
 These matrices are anti-symmetric ($ \alpha^{ab}_I= \alpha^{[ab]}_I$), {invariant under Lorentz transformations}, and self-dual ($i ^\star \alpha^{ab}_I=\alpha^{ab}_I$) ---hence $\bar \alpha^{ab}_I$ is anti self-dual. As mentioned above, the equations for the potentials can be derived from the equations for the fields. The reverse  is also true. Therefore, either set of equations completely describes the theory. Field equations similar to $\alpha^{ab}_{I} \partial_{a} H^I_+=0$ have been written before in \cite{DowkerDowker1966, Weinberg1964};  our equations $\alpha^{ab}_{I} \partial_{a} H^I_+=0$ are also equivalent to  Maxwell's equations in spinorial language \cite{PenroseRindler1984}. 
   
 \item{\bf Relation between $\vec H_{\pm}$ and the field strength $\F_{ab}$}
 
From the field strength $\F$ and its dual $\Fs$, we define the self- and anti self-dual two-forms $\F_{\pm}=\frac{1}{\sqrt{2}}(\F\pm\, i \, \Fs)$, that satisfy \ $ i \, ^\star \F_{\pm}=\pm \F_{\pm}$. The relation between the field strength and   $\vec H_{\pm}$ is then given by
\be  \label{FtoH} \F^{ab}_+=\alpha^{ab}_{I} H^I_+\, , \hspace{1cm}\F^{ab}_-=\bar \alpha^{ab}_{\dot I}  H^{\dot I}_- \, . \ee
These relations imply that one can understand the three $\alpha^{ab}_{I}$ matrices as a basis for the three-dimensional complex vector space of self-dual tensors in Minkowski spacetime (see Appendix \ref{The soldering form}). Then, $H^I_+$ are simply the components of $\F^{ab}_+$ in this basis. Similarly,  $\bar \alpha^{ab}_{I} $ provides a basis for anti self-dual tensors. 

On the other hand, by using the relations (\ref{FtoH}), and the fact that the $\alpha_{I}$-matrices are constant in spacetime, so they are transparent to derivatives, the field equations $\alpha^{ab}_{I} \partial_{a} H^I_+=0$ and  $\bar \alpha^{ab}_{I} \partial_{a} H^I_-=0$ can be written as $\partial_{a} \F^{ab}_+=0$ and $\partial_{a} \F^{ab}_-=0$, which are equivalent Maxwell's equations in their more standard form.

 \item{\bf Properties of the $\alpha^{ab}_{I} $ matrices}
 
 Using the form of the $\alpha_{I}$ matrices (\ref{alphas}), it is straightforward to check that they have the following properties
 
  \begin{itemize}
 \item Anti-commutation relations: $\{\alpha_I,\alpha_J\}\equiv \alpha^{a}_{\ b\, I} \, \alpha^{bc}_{\ \  J}+ \alpha^{a}_{\ b\, J}\,  \alpha^{bc}_{\ \ I}=\delta_{IJ} \, \eta^{ac}\, .$
 
 \item Commutation relations: $[\alpha_I,\alpha_J]\equiv \alpha^{a}_{\ b\, I} \, \alpha^{bc}_{\ \  J}- \alpha^{a}_{\ b\, J}\,  \alpha^{bc}_{\ \ I}= \, ^+\Sigma^{ac}_{\ \ IJ}\, .$
 
 \end{itemize}
 
 These properties  can be  thought as the spin-$1$ analog of the familiar properties of the Pauli matrices $\sigma^{A\dot A}_i$.

To better understand these properties, and to generalize them to curved spacetimes (see next section), it is convenient to take a more geometric viewpoint and think about the field $H^I_+$ as belonging to an abstract complex three-dimensional vector space $V$, that is support of a $(0,1)$ irreducible representation of the Lorentz group. This space is isomorphic to the space of self-dual tensors $\F_+$ in Minkowski spacetime, and  $\alpha^{ab}_I$ provides an isomorphism.

{Furthermore, $\alpha^{ab}_I$ equips $V$ with a product} $h_{IJ}$, the image of the Minkowski metric\footnote{I.e., given any two self-dual tensors $^{(1)}\F_+^{ab}$ and  $^{(2)}\F_+^{ab}$,  the isomorphism satisfies $^{(1)}\F_+^{ab}\, ^{(2)}\F_+^{cd} \, \eta_{ac}\, \eta_{bd}= \, ^{(1)} H^I_+\, ^{(2)}{H}^J_+ \,4 h_{IJ}$, where $ ^{(i)}\F_+^{ab}=\alpha^{ab}_{I} \  ^{(i)}H^I_+$ for $i=1,2$.} 
 $h_{IJ} =\frac{1}{4}\, \eta_{ab}\, \eta_{cd}\, \alpha^{ac}_I\, \alpha^{bd}_J$, whose value turns out to be $h_{IJ}= -\delta_{IJ}$ {in a cartesian frame}, and is obviously invariant under Lorentz transformations in $V$. 
This viewpoint makes clearer the analogy between  the  $\alpha^{a  b}_I$ and the Pauli  matrices $\sigma^{A\dot A}_i$ (recall that  $\sigma^{A \dot A}_i$ provides an isometry between spatial vectors and spinors). 
 
 If $\vec H_+$ is an element of the complex vector space $V$, then $\vec H_-$ is an element of $\bar V$, the complex conjugate space. Although naturally isomorphic, these two spaces are different, and  from now on we will use doted indices on elements of  $\bar V\ni  H^{\dot I}_-$.
The properties of $\bar \alpha^{ab}_{\dot I}$ are obtained by complex conjugating the properties of $\alpha^{ab}_I$ written above. The anti-commutation relations are identical. However, the conjugation changes the commutation relation to  $[\bar \alpha_{\dot I},\bar \alpha_{\dot J}]= \, ^-\Sigma^{ab}_{\ \ \dot I\dot J}$, where  now it is the generator of the $(1,0)$ representation of the Lorentz group that enters in the equation. 
Appendix \ref{The soldering form} contains further information about the properties of these tensors.

 \item{\bf Second order equations for the potentials $A^{+}_{a}$}
 
 We focus on $A^{+}_{a}$, since the derivation for  $A^{-}_{a}$ can be obtained from it by complex conjugation. The fastest way to obtain the familiar second order differential equation for $A^{+}_{a}$ is to  take time derivative of (\ref{AMeq}), use commutativity between spatial and time derivatives, and then use again (\ref{AMeq}) to eliminate the first time derivative  in favor of the curl. The result can then be written in  covariant form as $\Box A^{+}_{a}-\partial^{b}\partial_{a}A^{+}_{b}=0$. 

Alternatively,  we can use the following argument, that can be  straightforwardly generalized to curved spacetimes. Notice that the equations of motion $\bar \alpha^{ab}_{\dot I} \partial_{a} A^{+}_{b}=0$, imply that the two-form $ \partial_{[a} A^{+}_{b]}$ is  self-dual. This is because, on the one hand, the anti-symmetry of $\bar \alpha^{ab}_{\dot I}$ makes that only the anti-symmetric part of $\partial_{a} A^{+}_{b}$ contributes to the equations and, on the other,  because contraction with $\bar \alpha^{ab}_{\dot I}$ extracts the anti self-dual component of $ \partial_{[a} A^{+}_{b]}$. Therefore, when the equations of motion hold, $A^{+}_{a}$ is the potential of a self dual form, $\F_+ = \d A_+$. But if $\d A_+$ is self-dual, then the identity  $\partial_{[a} \partial_{b}A^{+}_{c]}=0$ implies that $\partial^{a} \partial_{[a}A^{+}_{c]}=0$.  These last equations are obviously equivalent to
\be  \Box A^{+}_{c}-\partial^{a}\partial_{c} A^{+}_{a}=0 \, .  \ee

Therefore, the self- and anti self-dual potentials $A^{\pm}_{a}$ satisfy the same second order equations than the ordinary vector potential.

\item{\bf Conserved current and charge}

In terms of self- and anti-self dual variables, electric-magnetic rotations take the simple  form
\be H^I_{\pm}(x) \rightarrow e^{\mp i\, \theta} \, H^I_{\pm}(x)\, , \hspace{0.75cm} A^{\pm}_{a}(x) \rightarrow e^{\mp i\,  \theta} \, A^{\pm}_{a}(x) \, .\ee
And the on-shell current  (\ref{osj}) takes the form 
\be j^{a}_D|_{\rm on-shell}=-\frac{i}{2} \Big[H_+^I\, \alpha^{ab}_{I}\, A^{-}_{b}-H_-^{\dot I}\, \bar \alpha^{ab}_{\dot I}\, A^{+}_{b}\Big]\, ,\ee
(note that this current is manifestly real). By using  the form of the generic solution to the field equations (\ref{solH}), we find that the conserved charge
\be Q_D=\int_{\mathbb R^3} \d^3x\,  j^{0}_D|_{\rm on-shell}=\int_{\mathbb R^3} \frac{\d^3k}{(2\pi)^3\, k}  \Big[ |h_+(\vec k)|^2-|h_-(\vec k)|^2\Big] \, , \ee
is proportional to the difference in the intensity of the self- and anti-self dual parts of field or, equivalently, the difference between the right and left circularly polarized components---i.e. the net helicity. ($Q_D$ has dimensions of angular momentum.) For this reason $Q_D$  is often called the {\em optical helicity}, or V-Stokes parameter. 

\end{enumerate}

\subsection{Curved spacetimes}  \label{Curved spacetimes}

The generalization  to curved spacetimes of the formalism just presented  follows the strategy commonly used for Dirac spin $1/2$ fields. Namely, one first introduces an orthonormal tetrad field, or Vierbein, in spacetime $e^{\mu}_a(x)$.\footnote{This non-coordinate orthonormal basis is defined by $g_{\mu\nu}(x)=\eta_{ab}\,  e^a_{\mu}(x)e^b_{\nu}(x)$, with $\eta_{ab}=diag\{+1,-1,-1,-1\}$. We assume our spacetime to admit such structure globally \cite{Geroch1968}.} 
With it, the curved spacetime $\alpha_I$-matices are obtained from the flat space ones $\alpha^{ab}_{I}$ by
\be \alpha^{\mu\nu}_{I}(x)=e^{\mu}_a(x)e^{\nu}_b(x)\, \alpha^{ab}_{I} \, . \ee
Furthermore, the Minkowski metric $\eta_{ab}$ is replaced by $g_{\mu\nu}(x)$; $\eta_{ab}$  is used to raise and lower flat-space indices $a,b,c,\cdots$,  $g_{\mu\nu}(x)$ for indices in the tangent space of the spacetime manifold $\mu,\nu,\beta,\cdots$, and $h_{IJ}$ and $h_{\dot I \dot J}$ for spin 1 indices. The matrices $ \alpha^{\mu\nu}_{I}(x)$ satisfy algebraic properties analog of the ones derived in Minkowski space
\be \{\alpha_I,\alpha_J\} \equiv \alpha^{\mu}_{\ \nu\, I} \, \alpha^{\nu \beta}_{\ \  J}+ \alpha^{\mu}_{\ \nu\, J}\,  \alpha^{\nu \beta}_{\ \ I}=-h_{IJ}\, g^{\mu \beta}\, ,  \ee \be \label{algprop} [\alpha_I,\alpha_J]\equiv \alpha^{\mu}_{\ \nu\, I} \, \alpha^{\nu \beta}_{\ \  J}- \alpha^{\mu}_{\ \nu\, J}\,  \alpha^{\nu \beta}_{\ \ I}= \, ^+\Sigma^{\mu \beta}_{\ \ IJ}\, , \ee 
where $^+\Sigma^{\mu \beta}_{\ \ IJ}=e_a^{\mu}e_b^{\beta}\, ^+\Sigma^{a b}_{\ \ IJ}$. The extension of the covariant derivative  $\nabla_{\mu}$  is obtained also by using standard arguments (see e.g. Appendix A of \cite{Ashtekar1991}). Namely, the action  of $\nabla_{\mu}$ on indices $I$ of fields $H_+^I\in V$ is uniquely determined by demanding compatibility with the isomorphism $\alpha^{\mu\nu}_{I}(x)$, $\nabla_{\beta}\alpha^{\mu\nu}_{I}(x)=0$ (see Appendix \ref{covder}). The result, as one would expect, agrees with the usual expression for the covariant derivative acting on fields of spin $s$ derived using group-theoretic methods, particularized to $s=1$
\bea \nabla_{\mu}\, H_+^I=\partial_{\mu}H_+^I-\frac12(w_{\mu})_{ab}\, ^+\Sigma^{ab\, I}_{\ \ \ J}\, H_+^J \, , \nonumber \\
\nabla_{\mu}\, H_-^{\dot I}=\partial_{\mu}H_-^{\dot I}- \frac12(w_{\mu})_{ab}\,  ^-\Sigma^{ab\, \dot I}_{\ \ \ \dot J}\, H_-^{\dot J}\,  , \eea
where $ ^{\pm}\Sigma$ are the generators of the $(0,1)$ and $(1,0)$ representations of the Lorentz algebra introduced in the previous section, and $(w_{\mu})_{ab}$ is the standard one-form spin-connection
\be (w_{\mu})^{a}_{\ b}=e^{a}_{\alpha}\partial_{\mu} e^{\alpha}_b+e^{a}_{\alpha}e^{\beta}_{b}\, \Gamma^{\alpha}_{\mu\beta} \, , \ee
with $\Gamma^{\alpha}_{\mu\beta}$  are the Christoffel symbols.

With this in hand, the generalization is straightforward:
\begin{enumerate}
\item{\bf Maxwell's equations for the fields}
\be \label{1oeqHcs}  \alpha^{\mu\nu}_{I} \nabla_{\mu} H^I_+=0\, , \ \ \ \bar \alpha^{\mu\nu}_{\dot I} \nabla_{\mu} H^{\dot I}_-=0\, . \ee
Note the similarity with Dirac's equation. The relation between $H_{\pm}$ and the self- and anti self-dual parts of the field strength $\F$ is  given by $\F_+^{\mu\nu}=\alpha^{\mu\nu}_{I}\, H^I_+$ and $\F_-^{\mu\nu}=\bar \alpha^{\mu\nu}_{\dot I} \, H^{\dot I}_+$. With this, and keeping in mind that $\nabla_{\mu}\alpha^{\beta\nu}_{I}(x)=0$, equations (\ref{1oeqHcs}) become $\nabla_{\mu}\F_+^{\mu\nu}=0=\nabla_{\mu}\F_-^{\mu\nu}$, {which is manifestly equivalent to Maxwell's equations $\nabla_{\mu} \F^{\mu\nu}=0$ by recalling that $\F_{\pm}=\frac{1}{\sqrt{2}}\left[\F\pm i^{\star}\F \right]$}

\item{\bf Potentials $A^{\pm}_{\mu}$}

The self- and anti self-dual potentials satisfy the first-order equations: 
\be \label{1oeqcs}  \bar \alpha^{\mu\nu}_{\dot I} \nabla_{\mu} A^{+}_{\nu}=0 \, ,  \hspace{1cm}   \alpha^{\mu\nu}_{I} \nabla_{\mu} A^{-}_{\nu}=0 \, .\ee
These are equivalent to Maxwell's equations. This can be easily seen by using the same argument we did in Minkowski spacetime, namely by noticing that, because $ \alpha^{\mu\nu}_{I}$ and $\bar \alpha^{\mu\nu}_{\dot I} $ project on self and anti self-dual forms, respectively,  these two equations are simply the self- and anti self-duality condition for the forms $\F_{+\, \mu\nu}\equiv 2  \nabla_{[\mu} A^{+}_{\nu]}$ and $\F_{-\, \mu\nu}\equiv 2  \nabla_{[\mu} A^{-}_{\nu]}$, respectively. This, in turns, implies that the identities $\d\F_{+}=0$, $\d\F_{-}=0$ are equivalent to Maxwell's equations $\nabla_{\mu}\F_{+}^{\mu\nu}=0$, $\nabla_{\mu}\F_{-}^{\mu\nu}=0$ (see footnote \ref{du} in appendix \ref{Maxwell equations in curved spacetime}).
Additionally, $\nabla_{\mu}\F_{\pm}^{\mu\nu}=0$ is equivalent to the second order equations $\nabla_{\mu} \nabla^{[\mu} A_{\pm}^{ \nu]}=0$. \\

The relation between $A^{\pm}_{\mu}$ and $H^I_{\pm}$ (before involving any equation of motion) requires of a foliation of spacetime in spatial Cauchy hyper-surfaces $\Sigma_t$, in the same way as the relation between the {electric and magnetic} fields and the standard vector potential does. Given the foliation associated with the definition of $\alpha_I^{\mu\nu}$, $A_{+\, \mu}$ and $H^I_{+}$ are related by means of  the  ``curl'':\footnote{Notice that this curl is independent of the connexion $\nabla_{\mu}$, due to the  antisymmetry of $\epsilon^{I\mu\nu}$ in $\mu$ and $\nu$. It is useful to keep this in mind in manipulating expressions involving $H^I_{\pm}$ and $A^{\pm}_{\mu}$.} 
\be \label{HvsA} H^I_+=i\, \epsilon^{I\mu\nu} \nabla_{\mu}A^{+}_{\nu} \, \ee
(and similarly for $A^{-}_{\mu}$ and $H^{\dot I}_{-}$) where $\epsilon^{I\mu\nu}$ is a `purely spatial' antisymmetric  mixed tensor (see Appendix \ref{3+1 spacetime decomposition} for its precise definition). As shown in Appendix \ref{Maxwell equations in curved spacetime}, one  can easily see that if $A^{+}_{\nu}$ is a solution of (\ref{1oeqcs}), then $H_+^I$ defined by (\ref{HvsA}) satisfies the field equations (\ref{1oeqHcs}). The reverse is also true.

\end{enumerate}
 
 \section{First order Lagrangian formalism: Dirac-type formulation} \label{First order Lagrangian formalism: Dirac-type formulation}
 
 The goal of this section is to write a Lagrangian for electrodynamics in terms of self- and anti self-dual variables. The similarity of equations (\ref{1oeqHcs}) and (\ref{1oeqcs}) with Dirac's equation
 motivates  {us to look for a first order Lagrangian (i.e.\ linear in time derivatives) and write it in a form that will make Maxwell's theory manifestly analog to Dirac's theory, where the mathematical structures associated with spin $s=1/2$ will be replaced by their $s=1$ analogs. This formulation will become very useful in the study of the electric-magnetic rotations in the quantum theory. 

 \subsection{First order Lagrangian} \label{Lagrangian framework}

Consider the  action
\bea \label{actionsd2} S[A_+,A_-] =  -\frac{1}{2}\int \d^4x\sqrt{-g}  \left[ H^{\dot I}_-\,  \bar \alpha^{\mu\nu}_{\ \ \dot I}\nabla_{\mu}A^{+}_{\nu} + H^{I}_+\,  \alpha^{\mu\nu}_{\ \ I}\nabla_{\mu}A^{-}_{\nu}  \right] \, .\eea
The Lagrangian density defined by the integrand  differs in a total derivative from the  standard Lagrangian $-\frac{1}{4}\sqrt{-g}\, \F_{\mu\nu}\F^{\mu\nu}$ ({after passing from first to second order formalism}), thus leading to the same dynamics. 
The independent variables in this action are $A_{\nu}^{\pm}$, and therefore $H^{I}_+$ and $H^{\dot I}_-$ are understood as shorthands for $i\, \epsilon^{I\mu\nu} \nabla_{\mu}A^{+}{\nu} $ and $-i\, \epsilon^{\dot I\mu\nu} \nabla_{\mu}A^{-}_{\nu}$, respectively. Note that this action is first-order in time derivatives of $A^{\pm}_{\mu}$, and second order in spatial derivatives. Extremizing the action with respect to $A^{+}_{\nu}$ produces the desired equations of motion {(see Appendix \ref{Deriving the equations of motion from the first-order action} for more details)} 
\be \label{djhkjh} \frac{\delta S}{\delta A^{+}_{\mu}}=0\hspace{0.5cm} \longrightarrow \hspace{0.5cm}  \bar \alpha^{\mu\nu}_{\dot I} \nabla_{\mu} H^{\dot I}_-=0\, ,\ee
and, as discussed above and proved in Appendix \ref{Maxwell equations in curved spacetime}, these last equations are equivalent to $\alpha^{\mu\nu}_{I} \nabla_{\mu} A^{-}_{\nu}=0$.  Similarly, from $\frac{\delta S}{\delta A^{-}_{\mu}}=0$ one obtains $\bar \alpha^{\mu\nu}_{\dot I} \nabla_{\mu} A^{+}_{\nu}=0$.

For the computations presented in the next section it is convenient to fix the Lorenz gauge, $\nabla_{\mu}A_{\pm}^{\mu}=0$. There is a remarkably simple way of incorporating this condition in the action (\ref{actionsd2}). All we need is to extend the domain of the indices $I$ and $\dot I$ from $\{1, 2, 3\}$ to $\{0, 1, 2, 3\}$, and define $\alpha^{\mu\nu}_0=\bar\alpha^{\mu\nu}_0\equiv -g^{\mu\nu}$. This is analog to  the familiar extension of the Pauli matrices $\vec \sigma$ by adding $\sigma^0$ (the identity), which conmutes with all $\sigma^i$, $i=1,2,3$.
{Algebraic properties of the $\alpha^{\mu\nu}_I$-matrices extended in this way appear in Appendix \ref{extalpha}.}

To simplify the notation, we will use the same name for the action and the tensors $ \alpha^{\mu\nu}_{\hspace{0.15cm} I}$, although from now on the index $I$ is understood to run from 0 to 3. The equations of motion still take  the same form
 \be \label{alphaMaxwellAcurvedlorenz } \bar \alpha^{\mu\nu}_{\hspace{0.15cm}\dot I}\  \nabla_{\mu} A^+_\nu=0 \ , \hspace{1cm}    \alpha^{\mu\nu}_{\hspace{0.15cm}I} \ \nabla_{\mu} A^-_\nu=0   \ , \ee  
but they now include the  Lorenz condition as the equation for $I=0$ ($\dot I=0$)
 \be \label{lorenz} g^{\mu\nu}  \nabla_{\mu} A^+_\nu=0 \ , \hspace{1cm}   g^{\mu\nu} \nabla_{\mu} A^-_\nu=0   \ . \ee  
Note that the action depends now on two new variables $H^0_{\pm}$, but they have the sole role of acting as Lagrange multipliers to enforce Lorenz's condition. 

Inspection of the action (\ref{actionsd2}) reveals that, contrary to the standard Maxwell's Lagrangian, the Lagrangian density in (\ref{actionsd2}) is manifestly invariant, $\delta \mathcal{L}=0$, under electric-magnetic rotations  $A^{\mu}_{\pm} \rightarrow e^{\mp i \, \theta} \, A^{\mu}_{\pm}$. It is now straightforward  to derive the Noether's current from it (see Appendix \ref{Deriving the Noether current from the first-order action})
\be j_D^{\mu}|_{\rm {on-shell}}=(-g)^{-1/2}\left(\frac{\delta \mathcal{L}}{\delta  \nabla_{\mu}  A_{+\, \nu}} \delta A_{+\, \nu}+\frac{\delta \mathcal{L}}{\delta \nabla_{\mu}  A_{-\, \nu}} \delta A_{-\, \nu}\right)|_{\rm on-shell} = \frac{i}{2} \Big[H^{\dot I}_-\,  \bar \alpha^{\mu\nu}_{\ \ \dot I}A_{\nu}^+ - H^{I}_+\,  \alpha^{\mu\nu}_{\ \ I}A_{\nu}^-\Big]\, .\ee
Using the relation between self- and anti self-dual variables, and ordinary variables $A_{\mu}$ and $\F^{\mu\nu}$, it is straightforward to check that this  expression agrees with  $j_D^{\mu}|_{\rm on-shell}$ obtained in section \ref{sec:2}, equation (\ref{osj}).

\subsection{Dirac-type Lagrangian} \label{Dirac-type Lagrangian}
The goal of this section is to re-write the action (\ref{actionsd2}) (including the Lorenz-gauge fixing term) in a more convenient form that will make the theory formally similar to Dirac's theory of spin $1/2$ fermions and will facilitate the computations in the next sections.

We first integrate by parts (\ref{actionsd2}), so $A^{\pm}$ and $H_{\pm}$ appear in a more symmetric form
\be \label{longS} S[A^+,A^-]=-\frac{1}{4} \int d^4x\sqrt{-g} \left[ H^{\dot I}_- \, \bar \alpha^{\mu\nu}_{\ \ \dot I}\,\nabla_{\mu}A_{\nu}^+ - A_{\nu}^+\bar \alpha^{\mu\nu}_{\ \ \dot I}\,\nabla_{\mu} H^{\dot I}_- + H^{I}_+ \,  \alpha^{\mu\nu}_{\ \ I}\nabla_{\mu}\, A_{\nu}^-  -   A_{\nu}^- \, \alpha^{\mu\nu}_{\ \ I}\nabla_{\mu} H^{I}_+ \right] \, . 
\ee
This action can now be written as
\be \label{DiracS} S[A^+,A^-]=-\frac{1}{4}\int d^4x\sqrt{-g}\ \bar \Psi \, i\beta^{\mu}\nabla_{\mu}\Psi \, \ee
where we have defined\footnote{We could have alternatively defined a couple of fields with two ``components", $\left( {\begin{array}{c}
 A^+ \\ H_- 
  \end{array} } \right)$ and $\left( {\begin{array}{c}
 A^- \\ H_+ 
  \end{array} } \right)$. Physical predictions would be obviously  the same, since we are just writing the same theory in different variables.  However,  the formal analogy with Dirac's theory is  cleaner if we use the four-``component" object $\Psi$ defined in (\ref{Psi}).} %
\bea \label{Psi}
\Psi=\left( {\begin{array}{c}
 A^+ \\ H_+  \\ A^-  \\ H_- \\
  \end{array} } \right)   \, , \hspace{.5cm} \bar \Psi =   ( A^+,   H_+,    A^-,   H_- ) \, , \hspace{.5cm} \beta^{\mu}= i\,     \left( {\begin{array}{cccc}
 0 & 0 & 0  & \bar\alpha^{\mu}  \\
  0  & 0 &  -\alpha^{\mu}  &0 \\
 0 &  \alpha^{\mu}  & 0  & 0 \\
- \bar  \alpha^{\mu}  & 0  & 0  &0 \\
  \end{array} } \right) 
\ . \eea 
{\bf Remark:}  It is convenient to include in the definition of $\Psi$ an arbitrary parameter $\ell^{-1}$ with dimensions of inverse of length multiplying $A^{\pm}$, and compensate it by  adding a global factor $\ell$ to the action.  The action remains  invariant, but the replacement $A^{\pm}\to \ell^{-1} A^{\pm}$ makes all the components of $\Psi$ and $\bar \Psi$ to have the same dimensions (namely, $\sqrt{{\rm energy}/{\rm length}^3}$). To simplify the notation, we will not write $\ell$ explicitly, but it should be taken into account in evaluating the dimensions of expressions containing $\Psi$ and $\bar \Psi$.\\

The exact position of the indices in the components of $\Psi$ and $\bar \Psi$ 
can be easily obtained by comparing (\ref{longS}), (\ref{DiracS}) and (\ref{Psi}). We have omitted them in the main body of this paper to simplify the notation, but  the details can be found in Appendix \ref{Definition of and and  properties}. Equation (\ref{DiracS}) is formally analog to the action of a Majorana 4-spinor describing a field with zero electric charge,  whose lower two components are complex conjugate from the upper ones.

From the  algebraic properties of the extended $\alpha$-matrices, (\ref{propiedad1ext}) and (\ref{propiedad3ext}), it is straightforward to check that $\beta^{\mu}$ satisfies the  Clifford algebra $\rm Cliff(3,1)$
\be \label{Cliff}\{\beta^{\mu},\beta^{\nu} \}=2g^{\mu\nu}\mathbb I \, . \ee
We also have that $\nabla_{\nu}\beta^{\mu}(x)=0$. These matrices can therefore be thought of as the spin $1$ analog of the Dirac $\gamma^{\mu}$ matrices.

We now define the ``chiral'' matrix
\bea \label{beta52}
\beta_5\equiv \frac{i}{4!}\epsilon_{\alpha\beta\gamma\delta}\beta^{\alpha}\beta^{\beta}\beta^{\gamma}\beta^{\delta} =  \left( {\begin{array}{cccc}
- \mathbb I &0  & 0  &0 \\
  0  & -\mathbb I & 0  & 0 \\
 0 & 0 & \mathbb I  & 0 \\
 0 & 0  &  0 & \mathbb I \\
  \end{array} } \right)
\eea
Some properties can be immediately checked out:
\bea
\{\beta^{\mu},\beta_5\}=0\, , \hspace{1cm} \beta_5^2=\mathbb I\,  
\ . \eea
 Further details and properties  can be found in  Appendix \ref{Definition of and and  properties}.

Although the  basic variables in the  action are the potentials $A^{\pm}_{\mu}$, at the practical level one can work by considering $ \Psi$ and $\bar \Psi$ as independent fields ---note that this is the same as one does when working with Majorana spinors. The equations of motion take the form

\be  \label{eqsPsi} \frac{\delta S}{\delta \bar \Psi }=0\hspace{0.5cm} \longrightarrow \hspace{0.5cm}i \beta^{\mu} \nabla_{\mu} \Psi=0 \, .\ee
They contain four equations, one for each of  the four components of $\Psi$. The upper two are the equations  $\bar \alpha^{\mu\nu}_{\hspace{0.15cm}\dot I}\  \nabla_{\mu} A^+_\nu=0$ and $\alpha^{\mu\nu}_{I} \nabla_{\mu} H^I_+=0$.  The lower two are complex conjugated equations. \\

Now, by acting on (\ref{eqsPsi}) witn $(-i \beta^{\alpha}\nabla_{\alpha})$ we obtain a second order equation for $\Psi$:
\be (-i \beta^{\alpha}\nabla_{\alpha})\,  i\beta^{\mu}\, \nabla_{\mu}\Psi=( \beta^{(\alpha}\,  \beta^{\mu)}+ \beta^{[\alpha}\,  \beta^{\mu]})\,\nabla_{\alpha}\nabla_{\mu} \Psi=(\Box+\mathcal{Q})\, \Psi=0\, \ee
where we have used (\ref{Cliff}) and defined 
\be \label{Q} \mathcal{Q}\, \Psi\equiv  \frac{1}{2}\beta^{[\alpha}\,  \beta^{\mu]}\, W_{\alpha\mu} \, \Psi\, \ee
with 
\be  \label{W} W_{\alpha\mu} \Psi \equiv [\nabla_{\alpha},\nabla_{\mu}]\Psi = \frac{1}{2}\, R_{\alpha\mu\sigma\rho}\,  \left( {\begin{array}{cccc}
  \Sigma^{\sigma\rho} & 0 & 0  & 0  \\
  0  & \,  ^+\Sigma^{\sigma\rho} & 0  &0 \\
 0 & 0  &  \Sigma^{\sigma\rho}  & 0 \\
0  & 0  & 0  &  \, ^-\Sigma^{\sigma\rho} \\
  \end{array} } \right)\Psi \,  .\ee
where $ \Sigma^{\sigma\rho}_{\ \  \alpha\beta}=\delta^{\rho}_{\alpha}\delta^{\sigma}_{\beta} - \delta^{\rho}_{\beta}\delta^{\sigma}_{\alpha}$ is the generator of the $(1/2,1/2)$ (real) representation of the Lorentz group, while $^+\Sigma^{\sigma\rho}_{IJ}$ and  $^-\Sigma^{\sigma\rho}_{\dot I \dot J}$ are the generators of the $(0,1)\oplus (0,0)$ and $(1,0)\oplus (0,0)$  representations, respectively.

Looking at the expression for $W_{\alpha\mu} \Psi$ we see that it contains real  terms, $R_{\alpha\mu\sigma\rho} \Sigma^{\sigma\rho}$, as well as complex ones, $R_{\alpha\mu\sigma\rho} \, ^{\pm}\Sigma^{\sigma\rho}$. The real terms come from the action of covariant derivatives on $A_{\pm}^{\mu}$. Since $A_{\pm}^{\mu}$ are  vectors in spacetime, their covariant derivative includes a connexion associated to the  $(1/2,1/2)$ representation of the Lorentz group.\footnote{  This does not mean, however,  that $A_{\pm}^{\mu}$ transform according to the  $(1/2,1/2)$ representation of the Lorentz group;  they do only up to a gauge transformation \cite{Weinberg1964}. See \cite{Bender1968} for a more  precise account of this issue. } The complex terms in $W_{\mu\nu} \Psi$ originate from the $(0,1)$ and $(1,0)$ representations, to which $\vec H_{\pm}$ are associated with.

The Poisson brackets for $\Psi$ and $\bar \Psi$ can be easily derived from the canonical relations $\{A^+_{\mu},H_-^{\dot I}\}=\gamma^{\dot I}_{\mu} \delta(\vec{x}, \vec{x}')$, in an analogous way as usually done for Majorana spinors, with the difference that in the situation under consideration in this paper, the Poisson brackets  must be promoted to {\em commutation relations} in the quantum theory. If anti-commutators were rather used, one would find  the  quantum propagator to violate causality, as expected from the spin-statistics theorem. Therefore, in spite of the fermion-like appearance of the formulation used in this section, we are describing a theory of bosons.

\subsubsection{Axial current}
We now describe how the electric-magnetic {symmetry} and its associated conservation law look like in the language introduced in this section. By using the chiral matrix $\beta_5$, the transformation reads
\bea \label{chiraltransf}
\Psi \rightarrow e^{ {}  i\theta \beta_5}\Psi \, , \hspace{1cm} \bar\Psi \rightarrow \bar\Psi e^{ {} i \theta \beta_5}
\eea
Notice that this have the same form as a chiral transformation for fermions. Looking at the form of $\beta_5$ in equation (\ref{beta52}), it is clear that the upper two components of $\Psi$, i.e.\ $(A_+, H_+)$, represent the self-dual, or positive chirality part of the field, while the lower two components $(A_-, H_-)$ contain the anti self-dual, or the negative chiral part. The Lagrangian density (\ref{DiracS}) is manifestly invariant under these transformation, and in terms of $\Psi$ the conserved current reads
\bea\label{jDdirac}
j_D^{\mu}= \frac{1}{4} \bar \Psi \beta^{\mu}\beta_5 \Psi \ . 
\eea
The associated Noether charge  is 
\be
Q_D= \int_{\Sigma_t} \d\Sigma_\mu \,  j_D^\mu
= \frac{1}{4} \int_{\Sigma_t}  \d\Sigma_3  \, \bar \Psi \beta^{0}\beta_5 \Psi \,  ,\ee
where  $\d\Sigma_3$ is the volume element of a space-like Cauchy hypersuface $\Sigma_t$. This expression for   $Q_D$ is equivalent to the one  obtained in previous sections [see eqn.\ (\ref{Cargadual})].

\section{The quantum anomaly} \label{The quantum anomaly}

In this section we analyze whether  the  classical symmetry  under electric-magnetic rotations persists in the quantum theory. The most direct avenue to meet this goal is to compute the vacuum expectation value of the divergence $\nabla_{\mu} j_D^{\mu}$. {A non-vanishing result would imply that the vacuum expectation value of the charge $ Q_D$ is not a constant of motion.}
For the sake of clarity, we perform the calculation using two different methods. First, we provide a direct computation of  $\left<\nabla_{\mu} j_D^{\mu}\right>$, in which the ultraviolet divergences are identified and subtracted in a covariant and self-consistent way, and then we reproduce the same result using Fujikawa's approach to anomalies based on  path integrals. These two methods illuminate complementary aspects of the calculation.

\subsection{Direct computation} \label{Direct computation}

Both $j_D^\mu$ and $\nabla_{\mu}j_D^\mu$ are operators quadratic in fields, and therefore the computation of  their expectation values must include renormalization subtractions to eliminate  potential divergences:
\be \langle \nabla_{\mu} j_D^\mu\rangle_{\rm ren} =\langle \nabla_{\mu} j_D^\mu\rangle-\langle \nabla_{\mu} j_D^\mu\rangle_{\rm Ad(4)} \, .\ee
In this expression, $\langle \nabla_{\mu} j_D^\mu\rangle_{\rm Ad(4)} $ indicate renormalization terms of  fourth  adiabatic order that we will compute  using the DeWitt-Schwinger asymptotic  expansion. More precisely, this renormalization scheme works by writing $\langle \nabla_{\mu} j_D^\mu\rangle$ in terms of the Feymann two-point function  {$ S(x,x')=-i \langle T \Psi(x) \bar \Psi(x')\rangle$},  and then by replacing it by $[S(x,x')- S(x,x')_{\rm Ad(4)}]$, where $S(x,x')_{\rm Ad(4)}$ denotes the DeWitt-Schwinger subtractions up to fourth  adiabatic order, and then taking the limit $x\to x'$. 

A convenient way to regularize {potential  infrared divergences is by introducing a parameter $s> 0$ in the theory (that will be send to zero at the end of the calculation)   replacing the wave equation $D\Psi=0$ by $(D{+}s)  \Psi=0$, where $D\equiv i\beta^{\mu}\nabla_{\mu}$ \cite{MossToms2014}. With all this: 
\bea  \label{jM} \nabla_{\mu} j_D^\mu(x)& =& \nabla_{\mu} \left[\frac{1}{4} \bar \Psi (x)\beta^{\mu} \beta_5 \,\Psi(x)\right] =  \frac{-i}{4} \left[\bar \Psi (x)\overset{{}_{\leftarrow}}{D}  \, \beta_5 \,\Psi(x)- \bar \Psi(x) \beta_5   \overset{{}_{\rightarrow}}{D} \Psi(x) \right] \nonumber \\ &=& \lim_{\substack{s \to 0 \\ x \to x'}} \frac{-i}{2} \, s\,  \bar \Psi(x) \beta_5  \Psi(x')=\lim_{\substack{s \to 0 \\ x \to x'}} \frac{-i}{2} \,s\,  {\rm Tr}[ \beta_5  \Psi (x) \bar \Psi(x') ]\, ,\eea
where we have used $\{\beta^{\mu},\beta_5\}=0$. If we now make a  choice of vacuum state  $|0\rangle$, we obtain\footnote{    We choose $x^0>x'^0$ without loss of generality, so that  $T \Psi(x) \bar \Psi(x')= \Psi(x) \bar \Psi(x')$.      }
\be \label{formal} \langle \nabla_{\mu} j_D^\mu\rangle= \lim_{\substack{s\to 0 \\ x \to x'}} \, \frac{  1   }{2} \,  s\,  {\rm Tr}\Big[ \beta_5 \, S(x,x',s)\Big]\, .  \ee 
The renormalized expectation value is then  given by
\be \label{renj} \langle \nabla_{\mu} j_D^\mu\rangle_{\rm ren}= \lim_{\substack{s\to 0\\ x \to x'}} \, \frac{ 1 }{2} \,  s\,  {\rm Tr}\Big[ \beta_5 \, \Big(S(x,x',s)-S(x,x',s)_{\rm Ad(4)}\Big)\Big]\, . \ee 
In this expression, $S(x,x',s)$ contains the information about the vacuum state, while  the role of $S(x,x',s)_{\rm Ad(4)}$ is to remove the potential ultra-violet divergences, which are the same for all vacua. It  is convenient to write $S(x,x',s)_{\rm Ad(4)}=   \left[ (D-s) G(x,x',s)\right]_{\rm Ad(4)}   $ ($D$ acts on the $x$-argument), where\footnote{\label{footk} This expression for  $ G(x,x',s)_{\rm Ad(4)}$ is obtained by writing $G(x,x',s)_{\rm Ad}$   first in terms of its heat kernel $K(\tau,x,x')$, $G(x,x',s)_{\rm Ad}=i\, \hbar\, \Delta^{1/2}(x,x') \int_{0}^{\infty} d\tau \, e^{-i\, (\tau  s^2+\frac{\sigma(x,x')}{2\tau})}\, K(\tau,x,x')$, and then by using the  asymptotic expansion $K(\tau,x,x)\sim \frac{ -i }{16\pi^2} \sum_{k=0}^{\infty}(i\tau)^{k-2}\,E_k(x)$ for $\tau \to 0$ . See e.g.\ \cite{parker-toms} for further details. 
}
\bea  \label{GAd} G(x,x',s)\sim  \frac{\hbar \Delta^{1/2}(x,x')}{16\pi^2} \sum_{k=0}^{\infty} E_k(x,x')  \int_0^{\infty} d\tau \, e^{-i\, (\tau  s^2+\frac{\sigma(x,x')}{2 \tau})}\, (i\tau)^{(k-2)} \,  \label{asympG}\eea
where $\sigma(x,x')$  is half of the geodesic distance square between $x$ and $x'$, $\Delta^{1/2}(x,x')$ is the Van Vleck-Morette determinant, and the functions $E_k(x,x')$ are the DeWitt coefficients, which are geometric quantities, built from the metric and its first $2k^{th}$ derivatives. We will only need the value of these coefficients when $x=x'$.  For  manifolds without boundary they are \cite{Vassilevich2003, parker-toms}
\bea \label{E2}
E_0(x)&=&\mathbb I  \, ,\nonumber \\
E_1(x)&=&\frac{1}{6}R\, \mathbb I-\mathcal Q\, ,  \nonumber \\
E_2(x) & = & \left[\frac{1}{72}R^2-\frac{1}{180}R_{\mu\nu}R^{\mu\nu}+\frac{1}{180}R_{\alpha\beta\mu\nu}R^{\alpha\beta\mu\nu}- \frac{1}{30} \Box R  \right] \mathbb I  \nonumber\\
 & + & \frac{1}{12}W_{\mu\nu}W^{\mu\nu}+\frac{1}{2}\mathcal Q^2-\frac{1}{6}R \mathcal Q+\frac{1}{6}\Box \mathcal Q,
\nonumber \eea
where the  expression for $W_{\mu\nu}\equiv [\nabla_{\mu}, \nabla_{\nu}]$ and $\mathcal{Q}(x)$ were given in (\ref{W}) and (\ref{Q}), respectively. $R$, $R_{\mu\nu}$, and $R_{\alpha\beta\mu\nu}$ are the Ricci scalar, Ricci tensor, and Riemann curvature tensor. 

Because of the symmetry of the classical action, the contribution of $S(x,x',s)$ to (\ref{renj}) vanishes for all choices of vacuum state. Therefore, $\langle \nabla_{\mu} j_D^\mu\rangle_{\rm ren}$ arises entirely from the subtraction terms, $S(x,x',s)_{\rm Ad(4)}$.  
This implies that   $\langle \nabla_{\mu} j_D^\mu\rangle_{\rm ren}$ is {\em independent of the choice of vacuum}. Notice that  the same  occurs in the calculation of other anomalies, such as the fermionic chiral anomaly or the trace anomaly. 

It turns our that only the terms with $k=2$ in (\ref{GAd}) produce a non-vanishing contribution. Furthermore, we do not need to consider terms  involving derivatives of $E_2(x,x')$, since they involve five derivatives of the metric and hence are of fifth adiabatic order. Taking into account that
\bea \label{E2}
& & {\rm Tr}[\beta_5 E_2(x,x))]  =  i\, \frac{1}{3}\, R_{\alpha\beta\mu\nu} \, ^\star R^{\alpha\beta\mu\nu} 
  \eea
where $ ^\star R^{\alpha\beta\mu\nu}=\frac{1}{ 2} \epsilon^{\alpha\beta\sigma\rho}R_{\sigma\rho}^{\hspace{0.35cm}\mu\nu}$ is the dual of the Riemann tensor, equation (\ref{renj}) produces:
\be \label{anomaly} \langle \nabla_{\mu} j_D^\mu\rangle_{\rm ren}=-\frac{\hbar}{96\pi^2}\, R_{\alpha\beta\mu\nu} \, ^\star R^{\alpha\beta\mu\nu} \, . \ee
Appendix \ref{Explicit calculations of the electromagnetic duality anomaly} contains  details of the intermediate steps in this computation.  A few comments are in order now:
\begin{enumerate}
\item This result reveals that quantum fluctuations spoil the conservation of the axial current $j_D^\mu$, and break the classical symmetry under electric-magnetic (or chiral) transformations. 
\item The pseudo-scalar  $R_{\alpha\beta\mu\nu} \, ^\star R^{\alpha\beta\mu\nu} $ is known as the Chern-Pontryagin density (its integral across the  entire spacetime manifold is the Chern-Pontryagin  invariant).
\item It is important to notice the parallelism with the chiral anomaly for spin $1/2$ fermions. The computations in that case would be very similar, except that one would have to use structures associated with  spin $1/2$ fields, rather than spin $1$. That would  change only the numerical coefficient in (\ref{anomaly}).

\end{enumerate}

\subsection{Path integral formalism} \label{Path integral formalism}

The functional integral for the theory under consideration is\footnote{As usual, the inclusion of the Lorentz gauge introduces two ghost scalar fields. These fields  do contribute to certain observables, such as  the trace anomaly. However, one can check explicitly that they do not affect the computation of  $\langle \nabla_{\mu} j_D^{\mu}\rangle $. It is for this reason that we have not written their  contribution to the path integral.}
\be \label{pathintegral0} Z=\int D\bar \Psi\,  D\Psi \, e^{i/\hbar\, S[\Psi, \bar \Psi]}\, .
\ee

The strategy of Fujikawa's approach to the computation of anomalies using path integrals is the following. 
The generating functional $Z$ is invariant under the replacement $(\Psi,\bar \Psi)\to  (\Psi'=e^{ i \beta_5 \theta}\Psi,\bar \Psi'=\bar \Psi e^{ i \beta_5 \theta})$, since this is just a change of variables and  the path integral  remains invariant under such a change.  However, the two components of the integrand, the measure and the action, could change under the transformation. Noether's theorem---in the version in which one considers  the parameter of the transformation $\theta(x)$ to be  a spacetime function of compact support---tells us that $\delta S=-\int d^4x\sqrt{-g}\, \theta(x) \, \nabla_{\mu} j^{\mu}_D$. On the other hand, the integral measure $D\bar \Psi\, D\Psi$ could change by a non-trivial Jacobian, $D\bar \Psi\, D\Psi\to J\, D\bar \Psi'\, D\Psi'$. Then, the invariance of $Z$ implies that these two changes must compesnate each other, i.e.\   $J\cdot e^{-i/\hbar \int d^4x\sqrt{-g}\theta(x) \,{\langle}  \nabla_{\mu} j^{\mu}_D{\rangle} }$ must be equal to one. From this we see that  quantum anomalies appear for those classical symmetries that do not leave the measure of the path integral  invariant, i.e.\, $J\neq 1$. The value of  $\langle \nabla_{\mu} j^{\mu}_D\rangle$ can  then be determined from $J$. The goal of this section is to  compute these quantities. 

The Jacobian $J$ can be determined by using standard functional analysis techniques  applied to the wave operator $D^2$, where $D=  \beta^{\mu}\nabla_{\mu}$. Consider the space of square-integrable fields $\Psi(x)$ with respect to the  product $\langle \Psi_1,\Psi_2 \rangle =\alpha\, \int d^4x \sqrt{-g} \, \Psi_1^{\dagger} \,  \Psi_2$, (see Appendix \ref{Definition of and and  properties} for further details, particularly the discussion around  expressions (\ref{product})),
and $\alpha>0$ is an arbitrary real parameter with dimensions of inverse of action.\footnote{\label{alpha}It is introduced in order to make the product dimensionless and, although $\alpha=\hbar^{-1}$ would be a natural choice, we leave it unspecified to make manifest that physical observables are independent of  it; it cancels out in intermediate steps.} It terms of the original variables $A_{\pm}$ and $H_{\pm}$, the norm of $\Psi(x)$ reads  $\langle \Psi,\Psi \rangle =\alpha \int d^4x \sqrt{-g} \, [ 2 \, |A_+|^2+2\,  |H_+|^2]\geq 0$.

It is easy to check that the operator $D^2$ is self-adjoint with respect to the product $\langle \Psi_1,\Psi_2 \rangle$. The self-adjointness of $D^2$ guarantees the existence of an orthonormal basis $\{\Psi_n\}$ made of eigenfunctions, $D^2\Psi_n=\lambda_n^2\, \Psi_n$. We will denote by $a_n$ the components of a vector $\Psi$ in this basis. An electric-magnetic rotation $\Psi\to \Psi'=e^{i\theta  \beta_5}\Psi$ can be now expressed as a change of the components $a_n \to a'_n =\sum_m\, C_{nm}\, a_m$, with $C_{nm}=\langle  \Psi_n, e^{  i \theta \beta_5}\, \Psi_m\rangle $. With this, the Jacobian of the transformation reads
\be D\bar \Psi\, D\Psi\to J\, D\bar \Psi'\, D\Psi'\,  ,\hspace{0.5cm} {\rm with} \hspace{0.5cm}  J=({\rm det} \, C)^{2}=e^{2\,  {\rm Tr}\, [\ln C]}=e^{i 2   \sum_n \langle  \Psi_n, \beta_5 \, \theta\, \Psi_n\rangle }\, \ee
Then, the invariance of the path integral  implies that, quantum mechanically
 \bea \langle \nabla_{\mu}j_D^{\mu} \rangle_{\rm ren}=  2\, \hbar \,\alpha \sum_{n=0}^{\infty}\, \bar \Psi_n\, \beta_5\Psi_n \, .
\eea
To evaluate this  expression we use again the heat kernel approach.  The  kernel of the equation,  $D^2\Psi=0$ is \cite{parker-toms}\footnote{The  factor $-4$ appears as a consequence of the fact that the pair of spinor fields that are canonically conjugated are $\Psi$ and $\bar \Psi$ ---and not $\Psi$ and $\frac{\partial L}{\partial \partial_t\Psi}=-\frac{1}{4}\bar \Psi$.} 
\be \label{K} K(\tau,x,x')=-4 \alpha  \sum_{n=0}^{\infty}\,e^{-i\, \tau \lambda_n^2} \Psi_n(x)\, \bar \Psi_n(x') \, \ee
Then 
\bea \label{Njpi} \langle \nabla_{\mu}j_D^{\mu} \rangle_{\rm ren}=\frac{-1}{2}  \hbar \lim_{\tau\to 0} {\rm Tr}\, [\beta_5\, K(\tau,x,x)]=i\frac{\hbar}{32\pi^2} \, {\rm Tr}[\beta_5E_2]=-\frac{\hbar}{96\pi^2}\, R_{\alpha\beta\mu\nu} \, ^\star R^{\alpha\beta\mu\nu} \ \, . \eea
where in the second equality we have used the  expansion of $K(\tau,x,x')$ for $\tau\to 0$, written in Footnote \ref{footk}, and in the last equality we have used (\ref{E2}).\\ 

{\bf Remark:} Recall that the path integral produces   transition amplitudes for time-ordered products of operators between the ``in" and ``out'' vacuum. However,   the result for  $\langle \nabla_{\mu}j_D^{\mu} \rangle_{\rm ren}$ comes entirely from the asymptotic terms in the heat kernel,  which are the same for all vacua. Therefore, the result (\ref{Njpi}) agrees with the {\em expectation value} of $ \nabla_{\mu}j_D^{\mu} $ in any vacuum state.

\section{Conclusions}

The apparently {trivial}  invariance of the source-free Maxwell's equations under duality transformations $\F_{\mu\nu}\to \Fs_{\mu\nu}$  has {interesting} physical consequences. 
This mapping can be extended to a continuous `rotation' $\F_{\mu\nu}\to \cos \theta\,  \F_{\mu\nu} \, +  \sin \theta \, \Fs_{\mu\nu}$, which can be proven to be a symmetry of  Maxwell's action both in flat and curved spacetimes. Noether's theorem provides then the existence of a conserved current and  the associated constant of motion, which describes the polarization state of electromagnetic radiation. The main {goal} of this paper {was} to show that this conservation law {does not survive the quantization} in curved spacetimes, {and an anomaly arises in the form of (\ref{Njpi})}. 

To meet our goal, we have re-written Maxwell's theory by using self- and anti-self dual variables. These fields transform under irreducible representations of the Lorentz group, and describe the two chiral sectors of the theory. In this language, Maxwell's electric-magnetic rotations reduce to an ordinary chiral transformation, {which in the absence of  charges and currents becomes a} symmetry of the classical theory. In this sense, our result can be understood as the spin $1$ generalization of the spin $1/2$ chiral anomaly.

Although anomalies arise mathematically  as a consequence of  taming ultraviolet divergences via regularization and renormalization, they have low-energy  implications, as  stressed e.g.\ in \cite{tHooft1979}. To give some examples, in two-dimensional spacetimes the trace anomaly implies the Hawking effect \cite{ChristensenFulling1977}, and the fermionic axial anomaly is closely related to the Schwinger pair creation effect  \cite{Blaeretal1981}. Similarly,  the electric-magnetic duality anomaly found in this paper is expected to have interesting physical applications in astrophysics, cosmology and {condensed matter systems}. This paper has been devoted to lay out the details of theoretical  formalism underlaying the computation of this anomaly. A detailed analysis of its physical consequences will be the focus of  future publications.  In particular, we expect that gravitational dynamics will be able to produce net circular polarization on photons through asymmetric creation of right/left  quanta. 
Some preliminary ideas were summarized in \cite{AgullodelRioPepe2017b}, where applications related to gravitational collapse and mergers in astrophysics were suggested.    \\

{\it Acknowledgments}.   
This work was supported by the Grants No. FIS2014-57387-C3-1-P;  FIS2017-84440-C2-1-P, Project No. SEJI/2017/042 (Generalitat Valenciana), the COST action CA15117 (CANTATA), supported by COST (European Cooperation in Science and Technology), and NSF CAREER Grant No. PHY-1552603. AdR was supported by the Spanish Ministry of Education Ph.D. fellowship no. FPU13/04948, and acknowledges financial support provided under the ERC Consolidator Grant "Matter and strong-field gravity: New frontiers in Einstein's theory", no. MaGRaTh-$646597$, PI: V. Cardoso. We thank A. Ashtekar, E. Bianchi and J. Pulin for useful discussions.

\appendix 

\section{Noether current} \label{Noether current}

We provide here a few more details about the variation of the Lagrangian density (\ref{variacionlagrangiano}) under the infinitesimal transformation (\ref{dtA}). We obtain
\bea
\delta \mathcal L & =  & \frac{\partial  \mathcal L}{\partial  A_{\nu}}\delta A_{\nu}+\frac{\partial  \mathcal L}{\partial \nabla_{\mu} A_{\nu}}\delta\nabla_{\mu} A_{\nu}=-{\sqrt{-g}}F^{\mu\nu}\nabla_{\mu} \delta A_{\nu}=-{\sqrt{-g}}\, \F^{\mu\nu}\nabla_{\mu} Z_{\nu} \, .
\eea
The equality $^{\star}\F=\d Z+\G$ leads to $\F=-^{\star}\d Z-\Gs$. Then $\Gs_{\mu\nu}\G^{\mu\nu}=({\Fs}_{\mu\nu}-\d Z_{\mu\nu})( -(^{\star} \d Z)^{\mu\nu} - \F^{\mu\nu})=\d Z^{\mu\nu}{(^{\star} dZ)_{\mu\nu}}  -\F^{\mu\nu}{^{\star} \F_{\mu\nu}}+2 \d Z^{\mu\nu}{\F_{\mu\nu}} $, from which we get
\bea
\delta \mathcal L= -\sqrt{-g}\, \frac{1}{2}\nabla_{\mu}(A_{\nu} {^{\star}\F^{\mu\nu}}-Z_{\nu} {^{\star} \d Z^{\mu\nu}})-\frac{1}{4}\, \sqrt{-g}\,  {^{\star} \G}_{\mu\nu}\G^{\mu\nu}\, .
\eea
The last term is equal to the product of the electric and magnetic parts of $\G$ and, since the latter vanishes in  one frame, ${^{\star} \G}_{\mu\nu}\G^{\mu\nu}=0$ in any frame. Then   $\delta \mathcal L$ is the divergence of a current, $\delta \mathcal L=\sqrt{-g}\, \nabla_{\mu} h^{\mu}$, which implies that the action remains invariant. 

The Noether current is then given by
\bea \label{asa}
j_D^{\mu} & = & \frac{1}{\sqrt{-g}} \frac{\partial \mathcal L}{\partial\nabla_{\mu} A_{\nu}}\delta A_{\nu}-h^{\mu}=  \frac{ 1}{2}\,  \Big[  A_{\nu}{\Fs}^{\mu\nu}-Z_{\nu}\, 2\, \F^{\mu\nu}-Z_{\nu} \, (^\star\d Z)^{\mu\nu}\Big] \, ,  \label{jND0apendice}
\eea
which agrees with (\ref{jND0}) after using  $ \d Z=  \, ^{\star}\F+\G$. 
{Acting now with the derivative operator on (\ref{jND0apendice}), one finds 
\bea
\nabla_{\mu}j_D^{\mu} &= &   \frac{ 1}{2}\,  \Big[  \nabla_{\mu}A_{\nu}{\Fs}^{\mu\nu}-2\nabla_{\mu}Z_{\nu}\, \, \F^{\mu\nu}-2Z_{\nu} \, \nabla_{\mu} \F^{\mu\nu} -\nabla_{\mu}Z_{\nu}\, (^{\star}dZ)^{\mu\nu} \Big] \nonumber\\
 &= &  \frac{ 1}{2}\,  \Big[  \nabla_{\mu}A_{\nu}{\Fs}^{\mu\nu}-(\Fs_{\mu\nu}+\G_{\mu\nu})\, \, \F^{\mu\nu}-2Z_{\nu} \, \nabla_{\mu} \F^{\mu\nu} -\frac{1}{2} (\Fs_{\mu\nu}+\G_{\mu\nu})(-\F^{\mu\nu} +\Gs^{\mu\nu}) \Big] \nonumber\\
&   = &  -Z_{\nu} \, \nabla_{\mu} \F^{\mu\nu} \nonumber
\eea
(Bianchi identity was used in the first  equality) which vanishes on-shell.}


\section{  The  $\alpha^{ab}_I$ tensor } \label{The soldering form}

This appendix contains additional properties of the $\alpha^{ab}_I$ tensors used in the main body of this paper. The properties for the tensors $\bar \alpha^{ab}_I$ are obtained by complex conjugation.

\subsection{Definition and properties\label{app2.1} }

Let  $\{t^{a},x^{a},y^a,z^{a}\}$ be an inertial  coordinate frame of contra-variant vectors  in 4D Minkowski spacetime. Consider the  following set of complex, antisymmetric tensors 
\bea
 \alpha^{ab}_{1} & = & -2\, (t^{[a}   x^{b]}+i \, y^{[a} z^{b]})\, ,\label{Xmunu}\\
 \alpha^{ab}_{2} & = & -2\, (t^{[a}   y^{b]}+i \, z^{[a} x^{b]})\, ,\\
 \alpha^{ab}_{3} & = & -2\, (t^{[a}   z^{b]}+i\,  x^{[a} y^{b]})\, ,\label{Zmunu} 
\eea
where the square brackets indicate antisymmetrization of indices. It is straightforward to check that they are self-dual, i.e\, $i\, ^\star\alpha^{ab}_{I}\equiv i\, \frac{1}{2}\, \epsilon^{ab}_{\ \ cd}\, \alpha^{cd}_{I}= \alpha^{ab}_{I}$. These three tensors form an orthogonal 
basis in the space of self-dual (complex) tensors in Minkowski spacetime. Given any such tensor $\F^{ab}_+$, we can write it as
\bea \label{iso}
\F^{ab}_+= {H}^{I}_+ \alpha_{I}^{ab}  \label{F+2} \, ,
\eea
where ${H}^{I}_+$ indicate the components of ${\F}^{ab}_+$ in this basis. This last equation can alternatively read as follows. Let $V$ be a 3-dimensional complex vector space, made of vectors ${H}^{I}_+$.  
Let $\{X_I, Y_I, Z_I\}$ be a  basis of one-forms  in the dual space $V^*$.   Equation (\ref{iso}) tells us that $\alpha_{I}^{ab}$ is an isomorphism between $V$ and the space of self-dual tensors.  An isomorphism can be obtained by identifying basis:
\bea
\alpha^{ab}_I\equiv \alpha_1^{ab} \, X_I + \alpha_2^{ab} \, Y_I + \alpha_3^{ab} \,  Z_I\, . \label{defalphas}
\eea
This isomorphism  can be used to endow $V$ with a product $h_{IJ}=\frac{1}{4}\, \eta_{ab}\, \eta_{cd}\, \alpha^{ac}_I\, \alpha^{bd}_J$, that in the basis we started with has components equal to minus the Kronecker delta, $-\delta_{IJ}$. Spacetime indices $a,b,c,\cdots$ are raised and lowered with Minkowski metric $\eta_{ab}$, while ``internal'' indices $I,J,K,\cdots$ are raised and lowered  with $h_{IJ}$. 

We collect here some useful properties of the tensors $\alpha^{ab}_{I}$, which can be checked by direct computation:
\bea 
\alpha_{ab I} \, \alpha^{ab}_{\ \ J} &=& 4\, h_{IJ} \label{propiedad2}\\
\alpha_{ab}^{\ \ I} \,  \alpha_{cd I}   &=&  4 \, ^+P_{abcd} \label{propiedad1} \\
\alpha_{ab I} \, \bar \alpha^{ab}_{\ \ \dot J} &=& 0 \label{propiedad3/2} \\
\alpha^{a}_{\ b I} \, \alpha^{cb}_{ \ \ J} & = &  h_{IJ}{\eta}^{ac}- \left[^+\Sigma_{IJ}\right]^{ac} \label{propiedad3}
\eea
In property (\ref{propiedad1}), $^+P_{abcd}=\frac{1}{4}(\eta_{ac}\eta_{bd}-\eta_{ad}\eta_{bc}+i \epsilon_{abcd})$ is the projector on self-dual tensors in Minkowski spacetime, and $\, \left[^+\Sigma_{IJ}\right]^{ac}$ is the generator of the $(0,1)$ representation of the Lorentz group, whose explicit form is $\left[^+\Sigma_{IJ}\right]^{ab}=-i\epsilon_{IJK}\, \alpha^{ab\, K}$.  Recall that, according to our sign conventions, we have $\eta_{ab}=t_a t_b-x_a x_b-y_a y_b - z_a z_b$ and $\epsilon^{abcd}=-4!\,  t^{[a} x^b y^c z^{d]}$ in this basis.
On the other hand, taking the symmetric and antisymmetric parts of (\ref{propiedad3}) yields the ``commutation" and ``anti-commutation" properties of  $\alpha_I^{ab}$:  
\bea
\alpha^{[a}_{\ \ b I} \alpha^{c]b}_{\hspace{0.15cm} \, J}   & = &   - \left[^+\Sigma^{ac}\right]_{IJ}  \label{propiedad3a}\\
\alpha^{(a}_{\ \ b I} \alpha^{c)b}_{\hspace{0.15cm} \, J}  & = &  \eta^{ac}h_{IJ} \label{propiedad3b}
\eea

In a similar manner, the tensor
\bea
 \bar \alpha^{ab}_{\dot I}\equiv \bar \alpha_1^{ab} \, X_{\dot I} +\bar \alpha_2^{ab} \, Y_{\dot I} +\bar \alpha_3^{ab} \,  Z_{\dot I}\, , \label{defalphascomplejo}
\eea
provides an isomorphism between the vector space $\bar V$, complex conjugated of  $V$, and the space of anti self-dual tensors in Minkowski spacetime
\bea \label{isoc}
\F^{ab}_-= {H}^{\dot I}_- \bar \alpha_{\dot I}^{ab}   \, .
\eea
The analog of properties of (\ref{propiedad2})-(\ref{propiedad3}) hold, replacing  $^+P^{abcd}$ by the anti self-dual projector $^-P^{abcd}$, that is simply the complex conjugated of $^+P^{abcd}$,  and $\, \left[^+\Sigma_{IJ}\right]^{ac}$  by the generator of the $(1,0)$ representations $ \left[^-\Sigma_{\dot I \dot J}\right]^{ac}$ .\\

The generalization to curved spacetimes is straightforward. Given a field of Vierbeins  $e_a^{\mu}(x)$, i.e.\ a field of orthonormal basis of tangent vectors in the spacetime manifold $(M,g_{\mu\nu})$, the $\alpha^{\mu\nu}_{I}$ tensor is constructed from the Minkowski space one  $\alpha^{ab}_{I}$ by
\be \alpha^{\mu\nu}_{I}(x)=e^{\mu}_a(x)e^{\nu}_b(x)\, \alpha^{ab}_{I} \, . \ee
This makes obvious that the properties (\ref{propiedad2})-(\ref{propiedad3}) generalize to curved spacetimes by simply  replacing the tensors $\eta_{ab}$ and $\epsilon_{abcd}$ by their counterparts in curved geometries, $g_{\mu\nu}$ and $\epsilon_{\mu\nu\alpha\beta}$.

\subsection{Covariant derivative operator\label{covder}}
In this appendix we provide some details regarding the extension of the  action of the covariant derivative to indices $I,J,K,\cdots$. 

Recall that  the Vierbein $e_a^{\mu}(x)$ at a given point of the spacetime manifold $(M,g_{\mu\nu})$ provides an isometry between the tangent space at $x$ and Minkowski spacetime. The extension of the action of the covariant derivative $\nabla_{\mu}$  on ``internal'' indices $a,b,c,...$ is obtained by demanding  $\nabla_{\mu} e^{a}_{\nu}(x)=0$. This defines the 
connection 1-form $\omega_{\mu}$
\bea
\omega_{\mu}^{ab} =  e^{a}_{\nu}\partial_{\mu} e^{b\, \nu}+\Gamma^{\nu}_{\mu\alpha}e^{a}_{\nu} e^{\alpha\, b}\, ,
\eea
where $\Gamma^{\nu}_{\mu\alpha}$ are the Christoffel symbols. Recall that $\omega_{\mu}^{ab} $ is antisymmetric, $\omega_{\mu}^{ab}=\omega_{\mu}^{[ab]}$ (as a consequence of $\nabla_{\mu}g_{\alpha\beta}=0$). To further extend  the action of  $\nabla_{\mu} $ to the complex vector space $V$, we follow the standard strategy. Namely, by linearity the difference between  any two possible extensions is characterized by 
\bea
(\nabla_{\mu}-\bar \nabla_{\mu} )H_I=-C_{\mu\, I}^{\hspace{0.4cm}J}\, H_J \, , \hspace{0.5cm} H_I \in V^* \, .
\eea
If we chose $\bar\nabla_{\mu}$ to be the ordinary derivative associated to a system of coordinates, $\bar\nabla_{\mu}=\partial_{\mu}$, we see that there are as many derivative operators as mixed tensors $C_{\mu\, I}^{\hspace{0.4cm}J}$. The most natural condition to single out one of them is to demand that $\nabla_{\mu}$ annihilates the isomorphism $\alpha^{\alpha\beta}_I(x)$
\bea
0\equiv \nabla_{\mu} \alpha^{\alpha\beta}_I= \partial_{\mu}\alpha^{\alpha\beta}_I+\Gamma^{\alpha}_{\mu\rho}\alpha^{\rho\beta}_I+ \Gamma^{\beta}_{\mu\rho} \alpha^{\alpha\rho}_I-C_{\mu\, I}^{\hspace{0.4cm}J}\alpha^{\alpha\beta}_J \, .\nonumber
\eea
Using now that  $\alpha^{\alpha\beta}_{\ \  I}(x)= e^{\alpha}_a(x)e^{\beta}_b(x) \alpha^{ab}_{\ \  I}$, together with the properties of $\alpha^{ab}_{ I}$, we obtain from the previous equation the form of $C_{\mu\, I}^{\hspace{0.4cm}J}$ 
\bea
C_{\mu\, I}^{\hspace{0.4cm}J} & = & \frac{1}{2}e^a_{\alpha}(\partial_{\mu}e^{\alpha}_c)\alpha^J_{ab}\alpha_I^{cb}+ \frac{1}{2}\alpha_{\alpha\beta}^J \Gamma_{\mu\rho}^{\alpha}\alpha^{\rho\beta}_{I}  = \frac{1}{2}\alpha_{ab}^J \left[e^a_{\alpha}(\partial_{\mu}e^{\alpha}_c)+\Gamma^{\nu}_{\mu\alpha}e^{a}_{\nu} e^{\alpha}_c \right]  \alpha^{cb}_{I}  \nonumber \\
& = &   \frac{1}{2}\alpha_{ab}^J \alpha^{\hspace{0.3cm}b}_{I\, c}  \omega_{\mu}^{ac} = \frac{1}{2}\alpha_{\hspace{0.15cm}b}^{[a \hspace{0.25cm} J}\alpha^{c]b}_{\hspace{0.35cm}I} \,\omega_{\mu a c}  = \frac12\omega_{\mu}^{\hspace{0.15cm}ab} \left[ ^+\Sigma_{ab}\right]^{\hspace{0.15cm}J}_{I} \label{dercov}
\eea
where $\left[ ^+\Sigma_{ab}\right]^{\hspace{0.15cm}J}_{I}$ is the generator of the $(0,1)$ representation of the Lorentz group. Therefore, the covariant derivative acting on the field $H^{+}_{I}$ is given by
\be \label{CDH+2}\nabla_\mu H^+_I= \partial_\mu H^+_I- \frac12 \omega_\mu^{ab} \, [^+\Sigma_{ab}]_I^{\ \ J}H^+_J\ . \ee
 Using the curved-space version of property (\ref{propiedad1}), one concludes that the condition $ \nabla_{\mu} \alpha^{\alpha\beta}_I=0$ in turn leads to the  condition $ \nabla_{\mu} \alpha_{\alpha\beta}^I=0$ for the dual space, yielding
\be \nabla_\mu H_+^I= \partial_\mu H_+^I- \frac12 \omega_\mu^{ab} \, [^+\Sigma_{ab}]^I_{\ \ J}H_+^J\ . \ee

Further useful equalities can be found. Looking at property (\ref{propiedad2}) in curved space,  the above conditions imply that $\nabla_{\mu} h_{IJ} =0$. A similar derivation shows that the covariant derivative of the tensors  $h^{IJ}$ or $h^I_J=h^{IK}h_{KJ}$ also vanishes. Finally, the covariant derivative of the totally antisymmetric tensors $\epsilon_{IJK} $ is zero. This is readily seen by noting from (\ref{propiedad3}) that $\nabla_{\mu}{^{+}}\Sigma^{\alpha\beta}_{IJ}=0$. Recalling that $\left[^+\Sigma_{IJ}\right]^{\alpha\beta}=-i\epsilon_{IJK}\, \alpha^{\alpha\beta\, K}$, then one concludes $\nabla_{\mu}\epsilon_{IJK}=0$. 

{By complex-conjugating (\ref{CDH+2})}, we obtain 
\be \nabla_\mu H_-^{\dot I}= \partial_\mu H_-^{\dot I}-  \frac12\omega_\mu^{ab} \, [^-\Sigma_{ab}]^{\dot I}_{\ \ \dot J}H_-^{\dot J}\ . \ee
where $\left[ ^-\Sigma_{ab}\right]^{\hspace{0.15cm}\dot J}_{ \dot I}$ is the generator of the $(1,0)$ representation of the Lorentz group. The tensors $h_{\dot I \dot J}$, $h^{\dot I \dot J}$, $\delta^{\dot I}_{\dot J}$ and $\epsilon_{\dot I \dot J \dot K}$ are also annihilated by $\nabla_{\mu}$.\\

\subsection{3+1 spacetime decomposition}  \label{3+1 spacetime decomposition}

A globally hyperbolic spacetime can always be foliated by a one-parameter family of spatial hypersurfaces $\Sigma_t$, $M \simeq \mathbb R\times \Sigma_t$ {\cite{Wald1984}}. If we call  $n^{\mu}$ to the unit time-like vector field everywhere orthogonal to  $\Sigma_t$, then $h_{\mu\nu}=g_{\mu\nu}-n_{\mu}n_{\nu}$ is the  induced spatial metric on $\Sigma_t$.

We can now use the isomorphism  $\alpha^{\mu\nu}_{I}$ defined in  (\ref{defalphas}) to build the following mixed tensors: 
\begin{itemize}

\item $\gamma_I^{\mu} :=  n_{\nu}\alpha^{\mu\nu}_{\ \ I}$ provides an isomorphism between complex vectors in $V$ and  (spatial) vectors in the tangent space of $\Sigma_t$.

\item $\gamma_{\dot I}^{\mu} := n_{\nu}\bar \alpha^{\mu\nu}_{\ \ \dot I}$ is similar to the previous   map replacing $V$  by its complex conjugated space $\bar V$.
\item $\gamma_I^{\dot I} := \gamma_I^{\mu}\gamma^{\dot I}_{\mu}$ provides an isomorphism between $V$ and $\bar V$.
\item $\epsilon^{I\mu\nu}:=  \gamma^I_{\beta}\, n_{\alpha}\, \epsilon^{\alpha\beta\mu\nu}$ defines a  totally antisymmetric, `purely spatial' tensor with mixed indices. 
\end{itemize}
From the last definition one can derive an identity that will be useful in later calculations
\bea
i\, 2\, \epsilon^{I \mu\nu}=\alpha^{\mu\nu I}-\bar \alpha^{\mu\nu \dot J}\, \gamma^I_{\dot J} \, .  \label{epsilonalfas}
\eea

As we have already mentioned,  $\alpha^{\mu\nu}_{I}$ provides a one-to-one correspondence between self-dual tensors $\F_+^{\mu\nu}$ and  elements  $H^I_+\in V$. We can now also build an isomorphism  between self-dual tensors $\F_+^{\mu\nu}$ and purely spatial vectors in spacetime constructed as  $H^{\mu}_+\equiv n_{\nu}{\F}_+^{\mu\nu}$. Indeed:
\bea
H_+^{\mu}  & \equiv & n_{\nu}\F_+^{\mu\nu}  =  n_{\nu} \alpha^{\mu\nu}_{\hspace{0.15cm}I} H_+^I = \gamma_{I}^{\mu} H_+^I  \label{HrelacionH}
\, .\eea
 From the above definitions, and using (\ref{propiedad1}) and (\ref{propiedad3}), one can easily verify the following properties,
\bea
\gamma^{\nu}_I \gamma^{\beta\,I } & = &  \alpha^{\mu\nu}_I \, n_{\mu}  \alpha^{\rho\beta I}\, n_{\rho}= -n^{\nu}n^{\beta}+g^{\nu\beta}= h^{\nu\beta}\, ,  \label{doblegamma1}\\
\gamma^{\nu}_I \gamma_{\nu\, J } & = & \alpha^{\mu}_{\hspace{0.2cm}\nu\, I} \, n_{\mu}\,   \alpha^{\rho\nu\, J} \, n_{\rho} =   \alpha^{(\mu}_{\hspace{0.2cm}\nu\, I} \, n_{\mu}\,   \alpha^{\rho)\nu\, J} \, n_{\rho}   = h_{IJ}\, .\label{doblegamma2}
\eea
This shows that $\gamma^{\nu}_I $ provides indeed an isometry between spatial complex vectors in $\Sigma_t$ and elements of $V$.
Notice  that $\nabla_{\mu} \gamma^{\nu}_I\neq 0$, but the spatial derivative of $H_+^I$ satisfies $D_{\mu}H_+^{\mu}=D_I H^I_+$:
\bea
D_{\mu}H_+^{\mu} & = & h^{\mu\nu}\nabla_{\mu}H^+_{\nu}=h^{\mu\nu}\nabla_{\mu}(\gamma_{\nu}^I H_I^+)=\gamma^{\mu I}\nabla_{\mu}H_I^+ + h^{\mu\nu}(\nabla_{\mu}\gamma_{\nu}^I)H_I^+ \nonumber \\
 & =  & \gamma^{\mu I}\nabla_{\mu}H_I^+ + h^{\mu\nu}\alpha^{\sigma \ \ I}_{\ \nu} (n_{\mu}a_{\sigma}+K_{\mu\sigma})H_I^+= \gamma^{\mu I}\nabla_{\mu}H_I^+\equiv D_I H^I_+ \, ,\label{elecciongauge}
\eea
where we have used that $\nabla_{\mu}n_{\sigma}=n_{\mu}a_{\sigma}+K_{\mu\sigma}$, $a_{\sigma}$ is the 4-acceleration of the vector field $n_{\nu}$ and $K_{\mu\nu}=K_{(\mu\nu)}$ the  extrinsic curvature of the three-dimensional sub-manifold $\Sigma_t$. 
Furthermore, if  $D_{I}H_+^{I}=0$ then $H_+^I$ can be written as  the ``curl" of a complex potential,
$ 
H_{+}^{I} = i \epsilon^{I\mu\nu} \nabla_\mu A_\nu^{+}  \label{curl} \, .
$
Indeed,
\bea
D_I H^I_{+}&=&D_{\alpha} H_{+}^{\alpha} =i D_{\alpha} \epsilon^{\alpha \mu\nu}\nabla_{\mu} A_{\nu}^{+}=i\,  \epsilon^{\alpha \mu\nu}D_{\alpha}\nabla_{\mu} A_{\nu}^{+} =  i\,\epsilon^{\alpha \mu\nu}\nabla_{\alpha}\nabla_{\mu} A_{\nu}^{+}\nonumber \\ &\propto&   \epsilon^{\alpha \mu\nu}R_{\alpha \mu\nu\beta}A_{+}^{\beta}=0\, , \label{dh0}
\eea
where $R_{\alpha \mu\nu\beta}$ is the Riemann tensor, and we have used (\ref{elecciongauge}), (\ref{HrelacionH}) and (\ref{doblegamma1}).


\subsection{Extended $\alpha_{ab}^I$ and its algebraic properties}\label{extalpha}

This section provides some details regarding  the `extended $\alpha_I^{\mu\nu}$-tensors', defined by extending the range of the index $I$ to run from 0 to 3, and setting  $\alpha_0^{\mu\nu}=-g^{\mu\nu}$. 

We begin defining $\alpha$ in Minkowski spacetime. Let $\hat V\equiv V \oplus \mathbb C $, equipped with a Lorentzian flat metric $\eta_{IJ}$, be now our `internal' vector space, where the indices $I,J$  run from $0$ to $3$. The complex 3-dimensional vector space $V$, defined in Appendix \ref{app2.1}, is now a subspace of $\hat V$.  Let $n_I$ denote a unit time-like vector ($\eta_{IJ}n^I n^J=1$) orthogonal to the $V$ subspace (i.e.\ $n^I m^I\eta_{IJ}=0$, for all $m^J\in V^{}$ ). It spans a 1-dimensional vector space.  The metric tensor in $\hat V$ can be written as $\eta_{IJ}=n_I n_J+h_{IJ}$, where $h_{IJ}$ is the metric tensor in $V$ used in appendix \ref{app2.1}.
Let $X_I,Y_I,Z_I,n_I$ be an orthonormal basis of  $\hat V^*$, the dual space of $\hat V$, with $n_I=\eta_{IJ}n^J$. We define now the   `extended' tensor $\alpha_I^{\mu\nu}$  by extending expression (\ref{defalphas}) as follows
\bea
  \alpha^{ab}_{ I}\equiv  \alpha_1^{ab} \, X_{ I} + \alpha_2^{ab} \, Y_{ I} + \alpha_3^{ab} \,  Z_{ I}-\eta^{ab} \, n_I \, . \label{defalphasext}
\eea
Therefore, we have
\bea
\alpha^{ab}_I=\alpha^{ab}_J\, h^J_I -n_I \eta^{ab} \, ,\label{alfaextendidadesc}
\eea
where $\alpha^{ab}_J h^J_I$ (the projection of $\alpha^{ab}_I$ on $V$) is  simply the $\alpha^{ab}_I$ tensor used in the previous subsection, before extending the range of the indices $I,J,K,\cdots$. 

The tensor $\alpha^{ab}_I$ defined in (\ref{defalphasext}) maps vectors in $\hat V$  to tensors in Minkowski spacetime of the form
\be \alpha_I:\, H^I_+ \longrightarrow  \alpha^{ab}_I H^I_+= \F_+^{ab} - H^0_+\, \eta^{ab} \, . \ee
$\F_+^{ab}$ is an antisymmetric self-dual tensor that in Minkowski spacetime transforms under the $(0,1)$ irreducible representation of the Lorentz group, while $H^0_+$  is a scalar function.  {Thus, the extended tensors $\alpha^{ab}_I$ map vectors in $\hat V$ to  tensors in Minkowski spacetime that transform under Lorentz under the $(0,1)\oplus (0,0)$ representation.

The properties (\ref{propiedad2})-(\ref{propiedad3}) must be replaced by
\bea 
\alpha_{ab I} \, \alpha^{ab}_{\ \ J} &=& 4\eta_{IJ} \, , \label{propiedad2ext}\\
\alpha_{ab}^{\ \ I} \,  \alpha_{cd I}   &=&  4 \, ^+P^{abcd} +{\eta}^{ab}{\eta}^{cd}=4 \, ^-P^{acbd} +{\eta}^{ac}{\eta}^{bd}\label{propiedad1ext} \, ,\\
\alpha_{ab I} \, \bar \alpha^{ab}_{\ \ \dot J} &=& 4n_I n_{\dot J} \label{propiedad3/2}\, , \\
\alpha^{a}_{\ b I} \, \alpha^{cb}_{ \ \ J} & = &  \eta_{IJ}{\eta}^{ac}- \, ^+M_{IJ}^{ac} \label{propiedad3ext}\, ,
\eea
where $^+M_{IJ}^{ac}\equiv\,  ^+\Sigma_{IJ}^{ac} +2\, { \alpha^{ab}_{\ \ K}h^K_{(I} n_{J)} =-4 ^+P^{ac}_{\ \ \ bd}\gamma^{b}_{(I}\gamma^d_{J)}   }$\footnote{  {We shall use the same notation for $\gamma_I^{\mu} :=  n_{\nu}\alpha^{\mu\nu}_{\ \ I}$ with the extended alpha tensors, as well as the rest of mixed tensors in Appendix \ref{3+1 spacetime decomposition}. Its distinction is clear from the context.  } } , and ${}^+\Sigma_{IJ}^{ac} $ is the  generator of the $(0,1)\oplus (0,0)$  representation of the Lorentz group.\footnote{Note that we denote the generator of the $(0,1)\oplus (0,0)$ representation  with the same symbol as the $(0,1)$ generator that we used in the previous subsections; the $(0,1)\oplus (0,0)$ generator has more components than the   $(0,1)$ one, namely the components corresponding to $I$ or $J$ equal $0$; but these components are all equal zero, hence we find appropriate using the same name for the two generators.}  From (\ref{propiedad3ext}), we obtain the commutation and anti-commutation relations 
\bea
\alpha^{[a}_{\ \ b I} \alpha^{c]b}_{\hspace{0.15cm} \, J}   & = &   -\, ^+M^{ab}_{IJ} \, ,  \label{propiedad3a}\\
\alpha^{(a}_{\ \ b I} \alpha^{c)b}_{\hspace{0.15cm} \, J}  & = &  \eta^{ab}\eta_{IJ} \, .\label{propiedad3b}
\eea

The generalization of these properties to curved spacetimes is done, again, by using a Vierbein or orthonormal tetrad $e^{\mu}_a(x)$, to write the relation between the curved spacetime 
 $\alpha_{ I}$-matrices and the flat spacetime ones
\be \alpha^{\mu\nu}_{I}(x)=e^{\mu}_a(x)e^{\nu}_b(x)\, \alpha^{ab}_{I} \, . \ee
The covariant derivative acting on the extended indices $I,J,K,\cdots$ can be determined following the arguments  in appendix \ref{covder}, but now we demand that $\nabla_{\mu}$ annihilates the extended tensor  $\alpha^{\mu\nu}_{ I}$, $\nabla_{\alpha}\alpha^{\mu\nu}_{ I}=0$. The result is what one could expect, namely
\be \label{CDH+3}\nabla_\mu H^+_I= \partial_\mu H^+_I- \frac12  \omega_\mu^{ab} \, ^+\Sigma_{ab\, I}^{\ \ \  \, J}H^+_J\ . \ee
where $ ^+\Sigma_{ab\, I}^{\ \ \ \, J}$ is the generator of the $(0,1)\oplus (0,0)$ Lorentz-representation.
The properties of the conjugate tensors $\bar \alpha_{ab}^{\ \ \dot I}$ are obtained in a similar way.  

 We finish this appendix by deriving a few useful relations. First of all, 
from (\ref{propiedad2ext})   we obtain  $\nabla_{\mu}\eta_{IJ}=0$. Second, by acting with $g_{\mu\nu}\nabla_{\alpha}$ on equation (\ref{alfaextendidadesc}), and by using $\nabla_{\alpha} \alpha^{\mu\nu}_I=0$, we obtain $\nabla_{\rho} n_I=0$  (since $\eta_{IJ}=n_In_J+h_{IJ}$, we also conclude that $\nabla_{\mu}h_{IJ}=0$). Recalling (\ref{alfaextendidadesc}) again, this last property implies that the covariant derivative defined in this section also annihilates the projection of $\alpha^{\mu\nu}_I$ into $V$, namely $\nabla_{\alpha}( \alpha^{\mu\nu}_I\, h^I_J)=0$. 


\section{Maxwell equations in curved spacetime} \label{Maxwell equations in curved spacetime}

This appendix  shows the equivalence between the equations of motion for the potentials (\ref{1oeqcs}) and the fields (\ref{1oeqHcs}). \\

First, we show that the equation for the potential $\bar \alpha^{\mu\nu}_{\dot I} \nabla_{\mu} A_{+\, \nu}=0$ implies the equation for the field $\alpha^{\mu\nu}_{I} \nabla_{\mu} H^I_+=0$. (We focus on self-dual fields, the derivation for anti self-dual fields can be obtained by complex conjugation.) To prove this, notice first that using the identity (\ref{epsilonalfas}), the equation for the potential implies that $2\, i\, \epsilon^{I\alpha\beta}\nabla_{\alpha}A_{+\, \beta}=\alpha^{\mu\nu\, I}\nabla_{\mu}A_{+\, \nu}$. Then, recalling that $H^I_+\equiv i\epsilon^{I\mu\nu}\nabla_{\mu}A^+_{\nu}$, we see that when $A_{+\, \nu}$ satisfies the equations of motion the relation between the field and the potential can be re-written as  $H_+^I=\frac{1}{2}\alpha^{\mu\nu\, I}\nabla_{\mu}A_{+\, \nu}$. Acting now with $\alpha^{\delta\rho}_I\nabla_{\rho}$ we obtain
\be \alpha^{\delta\rho}_I\nabla_{\rho}H_+^I=\frac{1}{2}\,  \nabla_{\rho} \, \alpha^{\delta\rho}_I\, \alpha^{\mu\nu \, I}\nabla_{\mu}A_{+\, \nu}=-2\,  \nabla_{\rho} \, \nabla^{[\delta}A_{+}^{\rho]}=0\, .\ee
where we have used that $-4\alpha^{\delta\rho}_I\, \alpha^{\mu\nu \, I}$ is a projector on self-dual fields and, in the last equality, that all solutions of $\bar \alpha^{\mu\nu}_{\dot I} \nabla_{\mu} A_{+\, \nu}=0$ are also solutions of  the second-order equations $\nabla_{\rho} \, \nabla^{[\delta}A_{+}^{\rho]}=0$ (remember the discussion below eq. (\ref{1oeqcs})).

Next, we want to show the reverse, i.e.,  starting from $\alpha^{\mu\nu}_{I} \nabla_{\mu} H^I_+=0$, we want to show that there exists a potential $A_{+\, \nu}$, related to $H^I_+$ by $H^I_+=i\, \epsilon^{I\mu\nu} \nabla_{\mu}A_{+\, \nu}$, that satisfies $\bar \alpha^{\mu\nu}_{\dot I} \nabla_{\mu} A_{+\, \nu}=0$.

We begin by noticing that, the identity $\nabla_{\mu}\alpha^{\alpha\beta}_I=0$ allows us to write the field equations as $\nabla_{\mu}(\alpha^{\mu\nu}_{I}  H^I_+)=0$. Because $\alpha^{\mu\nu}_{I}  H^I_+$ is a self-dual tensor, this equation implies that the two-form defined by $\F_{+\, \mu\nu}\equiv \alpha_{\mu\nu\, I}  H^I_+$ is closed,\footnote{\label{du} Notice that for self or anti-self dual two-forms, $\nabla_{\mu} \,  {w}^{\mu\nu}=0$ if and only if $\nabla_{\mu} \,  {^{\star}w}^{\mu\nu}=0$, the latter formula being equivalent to $\d w=0$.} $\d \F_+=0$. This allows the introduction of a potential one-form $A_{+\, \mu}$, $\F_+= \d A_+$. Then, $\d A_+$ is self-dual; this is to say,  the contraction of $\bar \alpha^{\alpha\beta}_{\dot I}$ and $\d A_+$ vanishes. But this is precisely the equation of motion we are looking for, $\bar \alpha^{\alpha\beta}_{\dot I} \nabla_{\alpha}A_{+\beta}=0$. It only  remains to prove that $H_+^I$ and $A_{+\mu}$ are related by means of a ``curl". To see this, we notice that since $\F_{+\, \mu\nu}\equiv \alpha_{\mu\nu\, I}  H^I_+$, we have $ \alpha_{\mu\nu\, I}  H^I_+=2 \nabla_{[\mu}A_{+\nu]}$. Multiplying both sides by $\alpha^{\mu\nu  J}$ produces, $H^J_+=\frac{1}{2} \alpha^{\mu\nu  J} \nabla_{[\mu}A_{+\nu]}$. Using now the relation (\ref{epsilonalfas}) and the equation for $A_{+\beta}$, this relation reduces to $H^J_+=i\epsilon^{J\mu\nu}\nabla_{\mu}A_{+\nu}$, and this is what we wanted to prove.


\section{Deriving the equations of motion from the first-order action} \label{Deriving the equations of motion from the first-order action}

In this appendix we derive    the equation of motion from the first order action, described in section \ref{Lagrangian framework}. We begin with the derivation of equations (\ref{djhkjh}) from the action  (\ref{actionsd2}). Recall that in this action we have not introduced yet the Lorentz gauge, and the indices $I,J,\cdots$ and $\dot I,\dot J,\cdots$ take values $1, 2, 3$. 

From the form of the action (\ref{actionsd2})
\bea
 S_M[A_+,A_-]=-\frac{1}{2}\int d^4x\sqrt{-g} \left[ H^{\dot I}_- \bar \alpha^{\mu\nu}_{\ \ \dot I}\nabla_{\mu}A_{\nu}^+ + H^{I}_+ \alpha^{\mu\nu}_{\ \ I}\nabla_{\mu}A_{\nu}^- \right] \, ,\eea
we have
\bea
0=\frac{\delta S_M}{\delta A^+_{\nu}}=\frac{1}{2} \bar \alpha^{\mu\nu}_{\hspace{0.3cm}\dot I}\nabla_{\mu} H_-^{\dot I}+\nabla_{\mu}\frac{i}{2}\epsilon^{I\mu\nu} \alpha^{\alpha\beta}_{\hspace{0.3cm}I} \nabla_{\alpha}A_{\beta}^- \, .
\eea
We use now the identity (\ref{epsilonalfas}) 
[note that 
$\nabla_{\mu}\epsilon^{I\mu\nu}\neq 0$] and (\ref{propiedad1}) to write 
\bea
0 & = & \frac{1}{2} \bar \alpha^{\mu\nu}_{\hspace{0.3cm}\dot I}\nabla_{\mu} H_-^{\dot I}	+	\frac{1}{4}\alpha^{\mu\nu I} \alpha^{\alpha\beta}_{\hspace{0.3cm} I} \nabla_{\mu}\nabla_{\alpha}A_{\beta}^- - \frac{i}{2} \nabla_{\mu} \bar  \alpha^{\mu\nu  \dot I}\epsilon^{\alpha\beta}_{\hspace{0.3cm}  \dot I} \nabla_{\alpha}A_{\beta}^- -\frac{1}{4}\bar\alpha^{\mu\nu \dot I} \bar\alpha^{\alpha\beta}_{\hspace{0.3cm} \dot I} \nabla_{\mu}\nabla_{\alpha}A_{\beta}^- \nonumber\\
 & = & \frac{1}{2} \bar \alpha^{\mu\nu}_{\hspace{0.3cm}\dot I}\nabla_{\mu} H_-^{\dot I}	-\frac{i}{2} \nabla_{\mu}\bar  \alpha^{\mu\nu \dot I}\epsilon^{\alpha\beta}_{\hspace{0.3cm} \dot I} \nabla_{\alpha}A_{\beta}^-  + [^+ P^{\mu\nu}]^{\alpha\beta} \nabla_{\mu}\nabla_{\alpha}A_{\beta}^-    - [^-P^{\mu\nu}]^{\alpha\beta} \nabla_{\mu}\nabla_{\alpha}A_{\beta}^- 
\eea
Recalling that $H_{-}^{\dot I}=- i \epsilon^{\dot I \mu\nu}\nabla_{\mu}A_{\nu}^{\pm}$, and using the Bianchi identity $\epsilon^{abcd}R_{bcde}=0$, we get
\bea
0 &  = &  \bar \alpha^{\mu\nu}_{\hspace{0.3cm}\dot I}\nabla_{\mu} H_-^{\dot I} + \frac{i}{2}\epsilon^{\mu\nu\alpha\beta}\nabla_{\mu}\nabla_{\alpha}A_{\beta}^- \nonumber\\
 & = & \bar \alpha^{\mu\nu}_{\hspace{0.3cm}\dot I}\nabla_{\mu} H_-^{\dot I}  + \frac{i}{4}\epsilon^{\mu\nu\alpha\beta}R_{\mu\alpha\beta\sigma}A^{\sigma}_- \nonumber\\
 & = &  \bar \alpha^{\mu\nu}_{\hspace{0.3cm}\dot I}\nabla_{\mu} H_-^{\dot I} \, .\label{maxwellaux}
\eea
Finally, as showed in Appendix \ref{Maxwell equations in curved spacetime}, these equations are equivalent to   $\alpha^{\mu\nu}_{I} \nabla_{\mu} A^{-}_{\nu}=0$.
Similarly, by differentiating the action with respect  to $A_-$ we obtain   $ \alpha^{\mu\nu}_{\hspace{0.3cm} I}\nabla_{\mu} H_+^{I}=0$, that implies $\bar \alpha^{\mu\nu}_{\dot I} \nabla_{\mu} A^{+}_{\nu}=0$.

Now we derive again the equations of motion, but  starting from the action that incorporates the Lorenz gauge condition. As explained in section \ref{Lagrangian framework}, this gauge condition is incorporated by extending the range of the indices $I,J,\cdots$ and $\dot I,\dot J,\cdots$  to take values $0,1, 2, 3$,  by introducing Lagrange multipliers $H_{\pm}^0$, and defining $\alpha^{\mu\nu}_0=-g^{\mu\nu}$. In order to take advantage of the calculation done a few lines above, we will keep now the indices $I,J,\cdots$ and $\dot I,\dot J,\cdots$ running from 1 to 3, and write explicitly  the Lorenz-gauge fixing term in the action: 

\bea S[A_{\pm},H_{\pm}^0]= -\frac{1}{2}\int dx^4 \sqrt{-g} \,   \left[ H^{\dot I}_-\,  \bar \alpha^{\mu\nu}_{\ \ \dot I}\nabla_{\mu}A_{\nu}^+ + H^{I}_+\,  \alpha^{\mu\nu}_{\ \ I}\nabla_{\mu}A_{\nu}^-  -  H^0_- \nabla_{\mu}A_+^{\mu} -H^0_+ \nabla_{\mu}A_-^{\mu} \right]
\eea

Variation with respecto to $H_{\pm}^0$ provides the Lorenz-gauge condition: $\nabla_{\mu} A^{\mu}_{\pm}=0$. Variation with respecto to $A^+_{\nu}$ yields
\bea
0 = \frac{\delta S}{\delta A_{\nu}^+}=\bar \alpha^{\mu\nu}_{\dot I} \nabla_{\mu} H_-^{\dot I}-\frac{1}{2}\nabla^{\nu} H^0_- \label{ecmotionextendida}\, .
\eea
Let's first focus on the $0$-component of this equation with respect to the (arbitrary) space-time decomposition used to relate $H_{\pm}$ and $A^{\mu}_{\pm}$. This is done by contracting  (\ref{ecmotionextendida}) with the time-like vector $n_{\nu}$. The term involving $H_-^{\dot I}$ vanishes ($D_{\dot I}H_-^{\dot I}=0$ by construction, see (\ref{dh0})), and we obtain $n^{\mu}\nabla_{\mu}H^0_{-}\equiv \partial_t H^0_-=0$. On the other hand, acting with $\nabla_{\nu}$ on equation (\ref{ecmotionextendida}) we get $\Box H^0_-=0$; the term involving $H_-^{\dot I}$ again vanishes, because $\bar \alpha^{\mu\nu}_{\dot I}\nabla_{\nu} \nabla_{\mu} H_-^{\dot I}=0$.\footnote{This last formula can be checked as follows:
\bea
\bar \alpha^{\mu\nu}_{\dot I}\nabla_{\nu} \nabla_{\mu} H_-^{\dot I} &  = &  \nabla_{\mu}\nabla_{\nu} \bar \alpha^{\mu\nu}_{\dot I}H^{\dot I}_- = \nabla_{\mu}\nabla_{\nu} F_-^{\mu\nu} = \frac{1}{2}R_{\mu\nu\alpha}^{\hspace{0.6cm}\mu}F_-^{\alpha\nu}+ \frac{1}{2}R_{\mu\nu\alpha}^{\hspace{0.6cm}\nu}F_-^{\mu\alpha} =  \frac{1}{2} R_{\nu\alpha}F_-^{\alpha\nu} -  \frac{1}{2} R_{\mu\alpha}F_-^{\mu\alpha}=0\, .
\eea  }
Now, if both $\Box H^0_-=0$ and $\partial_t H^0_-=0$ hold, then $D_ID^I H_-^0 = 0$ holds. Choosing that $H^0_-$ vanishes at spatial infinity, one gets $H^0_-=0$.
With this, equations (\ref{ecmotionextendida}) reduce to $\bar \alpha^{\mu\nu}_{\dot I} \nabla_{\mu} H_-^{\dot I}=0$, which are the correct equations of motion. An identical reasoning can be applied for $H_+^0$. Following now the same arguments as in Appendix \ref{Maxwell equations in curved spacetime}, we can  write  the (extended) first-order equations of motion for the potentials as
\bea
\alpha^{\mu\nu}_{I} \nabla_{\mu} A^{-}_{\nu}=0 \, ,
\eea
with $I=0,1,2,3$, the 0 component being the Lorentz gauge-fixing.

 
\section{Deriving the Noether current from the first-order action} \label{Deriving the Noether current from the first-order action}


In this section we derive the Noether current in first-order formalism by working directly with the  variables $A_+$ and $A_-$ and the action functional (\ref{actionsd2}).

The variations of the Lagrangian density  $\mathcal{L}=\mathcal{L}[A_+,A_-]$ under an infinitesimal electric-magnetic rotation of the potentials, $\delta A_{\pm}=\mp i \delta\theta A_{\pm}$, produes
\bea
\delta \mathcal{L} & = & \frac{\partial \mathcal{L}}{\partial A^+_{\mu}}\delta A^+_{\mu}  +  \frac{\partial \mathcal{L}}{\partial \nabla_{\mu} A^+_{\nu}}\delta \nabla_{\mu} A^+_{\nu} + c.c. \nonumber\\
 & = & -\frac{1}{2} H_-^{\dot I}  \bar \alpha^{\mu\nu}_{\hspace{0.3cm}\dot I}  (-i\delta\theta) \nabla_{\mu} A^+_{\nu} -\frac{1}{2} i \epsilon^{I \mu\nu} \alpha^{\rho\sigma}_{\hspace{0.3cm}I}\nabla_{\rho} A^-_{\sigma} (-i \delta\theta) \nabla_{\mu} A_{\nu}^+ + c.c.  \nonumber\\
 & = & \frac{i \delta\theta}{2} H_-^{\dot I}  \bar \alpha^{\mu\nu}_{\hspace{0.3cm}\dot I}   \nabla_{\mu} A^+_{\nu} + \frac{i \delta\theta}{2} H^I_+ \alpha^{\rho\sigma}_{\hspace{0.3cm}I}\nabla_{\rho} A^-_{\sigma}  + c.c. = 0
\eea
We find that, unlike in second-order formalism, the duality rotation leaves the Lagrangian invariant. The Noether current is now constructed as
\bea \label{jotra}
j_D^{\mu} & = & \frac{\partial \mathcal{L} }{\partial \nabla_{\mu} A^+_{\nu} } \delta A_{\nu}^+ + c.c. \nonumber\\
 & = & -\frac{1}{2} H_-^{\dot I}   \bar \alpha^{\mu\nu}_{\hspace{0.3cm}\dot I} (-i\delta \theta) A_{\nu}^+  -  \frac{1}{2} i \epsilon^{I \mu\nu} \alpha^{\rho\sigma}_{\hspace{0.3cm}I}\nabla_{\rho} A^-_{\sigma} (-i \delta\theta)  A_{\nu}^+  +c.c. \nonumber \\
 & = &  \frac{i\delta \theta}{2} [H_-^{\dot I} \bar \alpha^{\mu\nu}_{\hspace{0.3cm} \dot I}    A^+_{\nu} - H_+^I   \alpha^{\mu\nu}_{\hspace{0.3cm}I}    A^-_{\nu} ] +\left[-\frac{\delta\theta}{2} \epsilon^{I \mu\nu} A_{\nu}^+  \alpha^{\rho\sigma}_{\hspace{0.3cm}I} \nabla_{\rho} A^-_{\sigma}   + c.c.\right]
\eea
This expressions agrees with the result obtained in section \ref{sec:2},  eq. (\ref{jND0}). { Note that the last term in (\ref{jotra}) does not contribute to the associated Noether charge, and is proportional to the equations of motion, vanishing on-shell.   It is not difficult to find that it  agrees exactly with the last term in (\ref{jND0}).}


\section{Definition of $\Psi$ and $\beta^{\mu}$ and  properties} \label{Definition of and and  properties}


In this appendix we define  the fields $\Psi$ introduced in section  \ref{Dirac-type Lagrangian}, as well as the matrices $\beta^{\mu}$ and $\beta_5$, and discuss their properties.

Given the complex potentials $A^{\pm}_{\mu}$ and the  self- and antiself-dual fields, $H_+^I$ and $H_-^{\dot I}$, we define the  object
\bea
\Psi=\left( {\begin{array}{c}
 A_{+\, \nu} \\ H_+^{I}  \\ A_{-}^{ \nu}  \\ H_{-\, \dot I} \label{psiapendice}
  \end{array} } \right) 
\eea
Note that all the four components of this object are related: $ A^-_{\mu}$ is the complex conjugate of  $A^+_{\mu}$; $H_+^{I}=i\, \epsilon^{I\mu\nu}\nabla_{\mu}A^+_{\nu}$; $H_-^{\dot I}$ is the conjugated of $H_+^{I}$. Therefore, $\Psi$ is the spin 1 analog of a Majorana spinor, whose upper and lower components are related by complex conjugation (Majorana fields represent real spinors with zero electric charge). 

If we denote by $X$ the vector space of all $\Psi$, we define now the linear  map $\beta^{\mu}: X \longrightarrow X$ by
 \be
\beta^{\mu}\Psi=i \left( {\begin{array}{c}
\bar \alpha^{\mu}_{\ \nu\, \dot I}H_-^{\dot I} \\ - \alpha^{\mu \ I}_{\ \nu} A_-^{\nu} \\  \alpha^{\mu\nu}_{ I} H_+^{I}  \\  -\bar \alpha^{\mu\nu}_{\dot I} A^+_{\nu}  
  \end{array} } \right) \, .
\ee
which is well defined for all $\Psi$. In matrix notation
\bea
 \beta^{\mu}= i\,     \left( {\begin{array}{cccc}
 0 & 0 & 0  &\bar \alpha^{\mu}_{\ \nu\, \dot I} \\
  0  & 0 &  - \alpha^{\mu \ I}_{\ \nu}  &0 \\
 0 &  \alpha^{\mu\nu}_{ I}  & 0  & 0 \\
-\bar \alpha^{\mu\nu}_{\dot I}  & 0  & 0  &0 \\
  \end{array} } \right)\, .
\eea
Define now the product of two $\beta^{\mu}$ as the composite operation, $\beta^{\mu}\beta^{\nu}: X \longrightarrow X$, defined by $(\beta^{\mu}\beta^{\nu})\Psi=\beta^{\mu}(\beta^{\nu}\Psi)$. This is  linear, and leads to 
\bea
 \beta^{\mu}\beta^{\nu}= \,     \left( {\begin{array}{cccc}
   \bar\alpha^{\mu \ \dot J}_{\ \alpha} \bar\alpha^{\nu \beta}_{\dot J}& 0 & 0  & 0  \\
  0  & \alpha^{\mu \ I}_{\ \alpha} \alpha^{\nu \alpha}_{  J} & 0 &0 \\
 0 & 0 & \alpha^{\mu \alpha}_{I } \alpha^{\nu \ I}_{\ \beta}  & 0 \\
0 & 0  & 0  &\bar \alpha^{\mu \alpha}_{ \dot I} \bar \alpha^{\nu \ \dot J}_{\ \alpha} \\
  \end{array} } \right)\, .
\eea 
Using the properties  (\ref{propiedad3a})-(\ref{propiedad3b}) of the $\alpha_I$ matrices, one can easily write the symmetric and anti-symmetric parts of this expression in $\mu$ and $\nu$, and in particular the symmetric part  produces the anti-commutation relations written in (\ref{Cliff}). 

Define the ``chiral" matrix $\beta_5$ as  a linear map $\beta_5: X \longrightarrow X$  by $\beta_5   \equiv    \frac{i}{4!}\epsilon_{\mu\nu\alpha\beta} \beta^{\mu} \beta^{\nu} \beta^{\alpha} \beta^{\beta}$. Manipulating this expression we obtain
\bea
\beta_5 & = & \frac{i}{4!}\epsilon_{\mu\nu\alpha\beta} \beta^{[\mu} \beta^{\nu]} \beta^{[\alpha} \beta^{\beta]}  \nonumber\\
& = &  \frac{i}{6} \epsilon_{\mu\nu\alpha\beta}\left( {\begin{array}{cccc}
 4 \left[^+P^{\mu\nu}\right]_{\sigma}^{\ \rho}  \left[^+P^{\alpha\beta}\right]_{\rho}^{\ \delta} & 0 & 0  &0 \\
  0  &\frac{1}{4}\left[^+M^{\mu\nu}\right]^{I}_{\ K}  \left[^+M^{\alpha\beta}\right]^{K}_{\hspace{0.15cm}J}& 0 &0 \\
 0 & 0  & 4 \left[^-P^{\mu\nu}\right]^{\sigma}_{\ \rho}  \left[^-P^{\alpha\beta}\right]^{\rho}_{\hspace{0.15cm}\delta}   & 0 \\
 0  & 0  & 0  &\frac{1}{4} \left[^-M^{\mu\nu}\right]_{\dot I}^{\ \dot K}  \left[^-M^{\alpha\beta}\right]_{\dot K}^{\hspace{0.15cm}\dot J} \\
  \end{array} } \right) \nonumber \\  
  &  = &   \left( {\begin{array}{cccc}
- g_{\sigma}^{\ \rho} & 0 & 0  &0 \\
  0  & -h^{I}_{\ J} & 0 &0 \\
 0 & 0  & g^{\sigma}_{\ \rho}   & 0 \\
 0  & 0  & 0  &h_{\dot I}^{\ \dot J} \\
  \end{array} } \right) \, .\eea
The tensor $^{\pm}P$ is the projector on (anti)self-dual tensors defined below (\ref{propiedad3}),  and the tensors $\left[^{\pm}M^{\alpha\beta}\right]$ were defined below equation (\ref{propiedad3ext}). In the above calculation, we have used the self-duality property, $\pm i ^{\star}P^{\pm}=P^{\pm}$. The map $\beta_5$ has the following properties
\bea
\beta_5^2=   \left( {\begin{array}{cccc}
 g_{\sigma}^{\ \rho} & 0 & 0  &0 \\
  0  & h^{I}_{\ J} & 0 &0 \\
 0 & 0  & g^{\sigma}_{\ \rho}   & 0 \\
 0  & 0  & 0  &h_{\dot I}^{\ \dot J} \\
  \end{array} } \right)   \, , \hspace{.5cm} \{\beta_5, \beta^{\mu}\} = 0 \, .
\eea
A duality transformation can be implemented by means of the linear operation $T_{\theta}: X \longrightarrow X$, with $T_{\theta}=e^{i\theta \beta_5}$, $\theta\in \mathbb R$.

Let $X^*$ be now the dual space, the space of  linear functionals over $X$. Given $\Psi\in X$ as in (\ref{psiapendice}), we  define $\bar \Psi\in X^*$ by 
\bea
\bar \Psi:=\left({\begin{array}{cccc}
 A_+^{\nu} & H^+_{I}  & A^-_{\nu}  & H_-^{\dot I}
  \end{array} }\right) \, . \label{tumbado}
\eea
The action functional (\ref{DiracS}) is thus a well-defined quantity.
To construct a Hilbert space from X, we need to endow it with an inner product. Note that, while the product $\bar \Psi \Psi \in \mathbb C$ is well-defined, it does not produce a positive real number. We can define a (positive-definite) inner product as follows

\be \label{product} \langle \Psi_1,\Psi_2 \rangle =\alpha\, \int d^4x \sqrt{-g} \, \bar \Psi_1 \,\delta \,  \Psi_2 \, ,\ee
where $\alpha$ is an arbitrary positive real constant with dimensions of inverse of action (see footnote \ref{alpha}), and     $\delta: X \longrightarrow X$ is a linear application defined by
\bea
\delta\Psi=\left( {\begin{array}{c}
 A^-_{\nu} \\ H_-^{I}  \\ A_+^{\nu}  \\ H^+_{\dot I}
  \end{array} } \right)  \label{elmezclado}
\eea
which in matrix notation reads 
\bea
\delta = \left( {\begin{array}{cccc}
 0& 0 & \delta^{\mu}_{\nu}  &0 \\
  0  & 0 & 0 & \gamma^I_{\dot I} \\
 \delta^{\nu}_{\mu}  & 0  & 0   & 0 \\
 0  & \gamma^{ I}_{\dot I}  & 0  &0 \\
  \end{array} } \right)
\eea
($\gamma_{I\dot I}$ was defined in Appendix \ref{3+1 spacetime decomposition}). This operation is useful since now  $\bar \Psi \delta \Psi \geq0$.  By expanding the fields as in (\ref{tumbado}) and (\ref{elmezclado}) one checks that expression (\ref{product}) is  real, and in particular $\left< \Psi_1, \Psi_2\right>=\left< \Psi_2, \Psi_1\right> $. Linearity with second variable is trivial. 

The analog of this product for Dirac field is commonly written simply as $\langle \Psi_1,\Psi_2 \rangle =\alpha\, \int d^4x \sqrt{-g} \, \Psi^{\dagger}_1  \,  \Psi_2$, where the matrix $\delta(=\gamma^0)$ is implicit in $ \Psi^{\dagger} \equiv \bar \Psi \delta$ to simplify the notation (see e.g. \cite{Fujikawa1980}). Note, however, that the presence of $\delta$ is required in order to make the operation well-defined regarding the position of indices. We  use the product (\ref{product}) in section \ref{Path integral formalism}.


\section{Details in the  calculation of the electromagnetic duality anomaly} \label{Explicit calculations of the electromagnetic duality anomaly}

This appendix provides  details of the intermediate steps summarized in section \ref{The quantum anomaly} regarding the computation of $\langle \nabla_{\mu} j_D^\mu\rangle_{\rm ren}$.  In that section we needed to compute
\be \label{loquehayquecalcular} \langle \nabla_{\mu} j_D^\mu\rangle= \lim_{\substack{s\to 0 \\ x \to x'}} \, \frac{ 1 }{2} \,  s\,  {\rm Tr}\Big[ \beta_5 \, S(x,x',s)\Big]\, .  \ee 
where $S(x,x',s)_{\rm Ad(4)}=   \left[ (D_x-s) G(x,x',s)\right]_{\rm Ad(4)} $, and with the asymptotic expansion in (\ref{asympG}). There will be no need of knowing explicitly the asymptotic expansion of $\Delta^{1/2}(x,x')$, $\sigma(x,x')$, and $E_k(x,x')$  in the short distance limit .

We shall show first that the derivative term, $D_x G(x,x',s)$, does not contribute to $\langle \nabla_{\mu} j_D^\mu\rangle$. From this contribution one only has to consider the $k=0,1$ terms in the sum (\ref{asympG}), since the term with $k=2$  is of adiabatic order five. The action of the derivative on $G(x,x',s)$ produces three contributions: one that goes with $\nabla_{\mu}^x \Delta^{1/2}(x,x')$, another with $\nabla_{\mu}^x \sigma(x,x')$, and  another with $\nabla_{\mu}^x E_k(x,x')$. The first two appear multiplied by ${\rm Tr}\{\beta^{\mu}\beta_5 E_k(x)\}$, and this quantity  vanishes for both $k=0,1$. Regarding the contribution of $\nabla_{\mu}^x E_k(x,x')$, it vanishes, because of the limit $s\to 0$. To see this, notice that for $k=0,1$, the factor $\nabla_{\mu}^x E_k(x,x')$  appears multiplied  in the sum  (\ref{asympG}) by the following contributions, respectively,
\bea
\int_0^{\infty} d\tau e^{-i\left (\tau s^2+\frac{\sigma(x,x')}{2\tau} \right)} (i \tau)^2  & = & \frac{2i}{\sigma(x,x')}+O(s^2) \\
\int_0^{\infty} d\tau e^{-i\left (\tau s^2+\frac{\sigma(x,x')}{2\tau} \right)} (i \tau)  & = & 2i \log s  +O(s^0)
\eea
so  the limit $s\to 0$ in (\ref{loquehayquecalcular})  vanishes.

We shall show now that the other term contributing to $S(x,x',s)_{\rm Ad(4)}$, $s \, G(x,x',s)_{\rm Ad(4)}$, only provides a non-zero result by means of the $k=2$ term in the asymptotic sum (\ref{asympG}). First notice that the limit $x\to x'$ can be  safely taken. On the other hand, higher values of $k$ in (\ref{asympG}) provide contributions of more than 4 derivatives of the metric to $S(x,x',s)$, so they are of higher adiabatic order. The $k=0$ case vanishes because it is proportional to ${\rm Tr}\{ \beta_5 E_0(x)\}={\rm Tr}\{\beta_5\}=0$. The $k=1$ term does not contribute either because its proportional to ${\rm Tr}\{ \beta_5 E_1(x)\}={\rm Tr}\{\beta_5 \mathcal Q \}$, and\footnote{In this calculation we used the relation $^+ P_{abcd}=-\frac{1}{4}\left[ \Sigma_{abcd}+i \left(^{\star}\Sigma_{ab}\right)_{cd} \right]$, and the Bianchi identity $R_{\mu\nu\alpha\beta}\epsilon^{\mu\nu\alpha\rho}=0$ (several times).}
\bea
\rm \rm Tr (\beta_5 \mathcal Q) & = & -2 i R_{\mu\nu\alpha\beta}\, \rm Tr \, Im[ ^+P^{\mu\nu} \Sigma^{\alpha\beta}- \frac14 \, ^+M^{\mu\nu}\, ^-\Sigma^{\alpha\beta}    ]  \nonumber\\
& = & \frac12 i R_{\mu\nu\alpha\beta}\, \rm Tr \, Im[  ^+M^{\mu\nu}\, ^-\Sigma^{\alpha\beta}    ]  \nonumber\\
& = & -2 i R_{\mu\nu I J} \epsilon^{\mu\nu \alpha I} = 2 i R_{\mu\nu \alpha \rho } \epsilon^{\mu\nu \alpha \sigma }n^{\rho}n_{\sigma}  = 0 \,   . \nonumber
\eea
Then, only remains to calculate the $k=2$ term in the asymptotic sum (\ref{asympG}),
\bea
\left< \nabla_{\mu}j_D^{\mu} \right> & = &  \frac{i \hbar}{32\pi^2} \rm Tr(\beta_5 E_2) \\ 
& = &   \frac{i \hbar}{32\pi^2}\left[\frac{1}{12} \rm Tr(\beta_5 W_{\mu\nu}W^{\mu\nu})+\frac{1}{2}\rm Tr(\beta_5 \mathcal Q^2) \right] \nonumber
\eea
with  $W_{\mu\nu}\equiv [\nabla_{\mu}, \nabla_{\nu}]$  given in (\ref{W}) and 
\be \label{QA} \mathcal{Q}\, \Psi\equiv  \frac{1}{2}\beta^{[\alpha}\,  \beta^{\mu]}\, W_{\alpha\mu} \, \Psi=-\frac{1}{2}R_{\mu\nu\alpha\beta} \left( {\begin{array}{cccc}
 -2 ^+P^{\mu\nu} \Sigma^{\alpha\beta} &0&0&0 \\
 0&  \frac12 ^+M^{\mu\nu}\, ^+\Sigma^{\alpha\beta} & 0&0  \\
0 & 0& -2  ^-P^{\mu\nu}  \Sigma^{\alpha\beta} &0\\
0&0&0& \frac12 ^-M^{\mu\nu}\, ^-\Sigma^{\alpha\beta} \\
 \end{array} } \right)\Psi \nonumber 
\, \ee
where $\Sigma^{\alpha\beta}_{\ \ \mu\nu}$ is the generator of the $(1/2,1/2)$ representation of the Lorentz group, $^+\Sigma^{\alpha\beta}_{\ \ IJ}$ is the generator of the $(0,1)\oplus(0,0)$ representation, and $^-\Sigma^{\alpha\beta}_{\ \ \dot I \dot J}$ of the $(1,0)\oplus(0,0)$ one, and $^{\pm}P_{abcd}=\frac{1}{4}(\eta_{ac}\eta_{bd}-\eta_{ad}\eta_{bc}\pm i \epsilon_{abcd})$.
A lengthy but straightforward computation produces
\bea
{\rm Tr}(\beta_5W_{\mu\nu}W^{\mu\nu}) & = & -2  i R_{\mu\nu \alpha\beta }{^{\star}R}^{\mu\nu\alpha\beta}
\eea
\bea
{\rm Tr}(\beta_5 \mathcal Q^2)  & = & i R^{\mu\nu \alpha\beta}\, ^{\star}R_{\mu\nu \alpha\beta} 
\eea
With this, we obtain
\bea
{\rm Tr}(\beta_5 E_2)= i  \frac{1}{3} \, R^{\mu\nu \alpha\beta}\, ^{\star}R_{\mu\nu \alpha\beta} \, . \eea

\bibliographystyle{unsrt}
 \bibliography{ELECTROMAGNETIC-DUALITY-ANOMALY}

\end{document}